\documentclass{article}
\usepackage[style=mla]{biblatex}
    \DeclareLanguageMapping{english}{english-apa}
\usepackage[letterpaper,top=2cm,bottom=2cm,left=2.5cm,right=2.5cm,marginparwidth=1.75cm]{geometry}
\usepackage[colorlinks=true, allcolors=blue]{hyperref}
\usepackage{authblk}

\usepackage{graphicx}
    \graphicspath{{./figures/}}
\usepackage{tabularx}
\usepackage{multirow}
\usepackage{makecell}
\usepackage{booktabs}
\usepackage{threeparttable}
\usepackage{float}
\usepackage{amsmath}
\usepackage{siunitx}
\usepackage[linesnumbered,ruled,vlined]{algorithm2e}
\usepackage{algpseudocode}

\usepackage{caption}
\usepackage{cleveref}
    \crefname{section}{Section}{Sections}
    \DeclareCaptionType{sfigure}[Supplementary Figure][Supplementary Figures]
    \DeclareCaptionType{stable}[Supplementary Table][Supplementary Tables]
    \DeclareCaptionType{sequation}[Supplementary Equation][Supplementary Equations]
    \crefname{sfigure}{Supplementary Fig.}{Supplementary Figs.}
    \Crefname{sfigure}{Supplementary Figure}{Supplementary Figures}
    \crefname{figure}{Fig.}{Figs.}
    \Crefname{figure}{Figure}{Figures}
    \crefname{stable}{Supplementary Table}{Supplementary Tables}
    \Crefname{stable}{Supplementary Table}{Supplementary Tables}
    \crefname{table}{Table}{Tables}
    \Crefname{table}{Table}{Tables}
    \crefname{sequation}{Supplementary Eq.}{Supplementary Eqs.}
    \Crefname{sequation}{Supplementary Equation}{Supplementary Equations}
    \crefname{equation}{Eq.}{Eqs.}
    \Crefname{equation}{Equation}{Equations}
\author[,1,3]{Helen H. Shang, MD, MS\thanks{These authors contributed equally to this work.}\hspace{0.5mm}}
\author[*,3,4]{Mohammad Sadegh Nasr\hspace{0.5mm}}
\author[3,4]{Jai Prakash Veerla\hspace{0.5mm}}
\author[3,4]{Jillur Rahman Saurav\hspace{0.5mm}}
\author[3,4]{Amir Hajighasemi\hspace{0.5mm}}
\author[3,4]{\mbox{Parisa Boodaghi Malidarreh\hspace{0.5mm}}}
\author[3]{Manfred Huber, PhD \hspace{0.5mm}}
\author[2]{Chace Moleta, MD\hspace{0.5mm}}
\author[2]{Jitin Makker, MD\hspace{0.5mm}}
\author[,3,4,5,6]{\mbox{Jacob M. Luber, PhD\thanks{Corresponding author. Email: \href{mailto:jacob.luber@uta.edu}{jacob.luber@uta.edu}}\hspace{0.5mm}}}

\affil[1]{Department of Internal Medicine, Ronald Reagan University of California Los Angeles Medical Center}
\affil[2]{Department of Pathology \& Laboratory Medicine, Ronald Reagan University of California Los Angeles Medical Center}
\affil[3]{Department of Computer Science and Engineering, The University of Texas at Arlington}
\affil[4]{Multi-Interprofessional Center for Health Informatics, The University of Texas at Arlington}
\affil[5]{Department of Bioengineering, The University of Texas at Arlington}
\affil[5]{Department of Biology, The University of Texas at Arlington}
\title{\Large\textbf{Histopathology Slide Indexing and Search: Are We There Yet?}}

\date{}
\begin{document}
    \maketitle
    
    \begin{abstract}

    The search and retrieval of digital histopathology slides is an important task that has yet to be solved. In this case study, we investigate the clinical readiness of four state-of-the-art histopathology slide search engines, Yottixel, SISH, RetCCL, and HSHR on both unseen datasets and several patient cases. We provide a qualitative and quantitative assessment of each model's performance in providing retrieval results that are reliable and useful to pathologists. We found high levels of performance across all models using conventional metrics for tissue and subtyping search. Upon testing the models on real patient cases, we found the results were still less than ideal for clinical use. Based on our findings, we propose a minimal set of requirements to further advance the development of accurate and reliable histopathology image search engines for successful clinical adoption. 

\end{abstract}
    \section{Introduction}

\begin{figure}[!t]
    \centering
    \includegraphics[width=\textwidth]{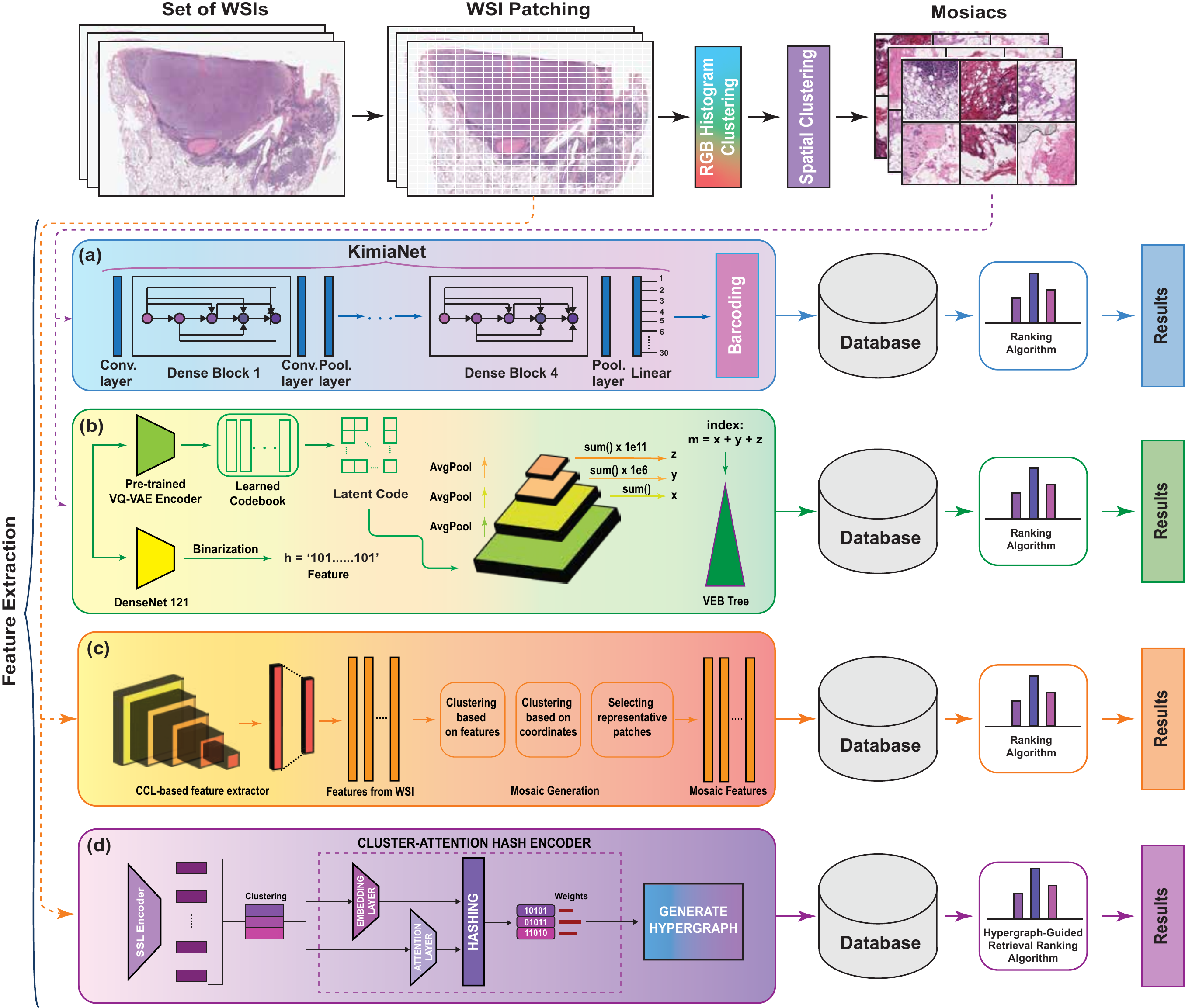}
    \caption{A summary of feature extraction and database creation processes proposed by \textbf{(a)} Yottixel \parencite{kalra_yottixel_2020}, \textbf{(b)} SISH \parencite{chen_fast_2022}, \textbf{(c)} RetCCL \parencite{wang_retccl_2023}, and \textbf{(d)} HSHR \parencite{li_high-order_2023}. The feature extractor of Yottixel is switched with KimiaNet \parencite{riasatian_fine-tuning_2021}.}
    \label{fig:summary_of_methods}
\end{figure}



    As histopathology slides become increasingly digitized, the process of manually searching and retrieving slides has become increasingly more time-consuming for pathologists \parencite{hegde_similar_2019, li_large-scale_2018}. Recently, there has been growing interest in the development of automated search and retrieval systems for digital histopathology slides \parencite{kalra_yottixel_2020, chen_fast_2022, wang_retccl_2023, hegde_similar_2019, kalra_pan-cancer_2020, li_high-order_2023}, which can help pathologists identify similar cases in both developing and narrowing a differential diagnosis. These systems leverage advances in artificial intelligence and machine learning to analyze large volumes of slides efficiently and accurately. 

    The exploration of medical image databases predominantly relies on content-based image retrieval (CBIR) \parencite{kalra_yottixel_2020, li_large-scale_2018, lew_content-based_2006}. CBIR systems initially transform images into a feature-based database accompanied by corresponding indices. Subsequently, by utilizing a similarity metric, the retrieval process simplifies into a k-nearest neighbors problem. Extracting features from extensive whole slide images (WSI) is typically achieved through either the sub-setting method, which focuses on a small section of a large pathology image to significantly reduce processing time, or the tiling method, which segments images into manageable patches (i.e., tiles) for intra-patch processing \parencite{kalra_yottixel_2020, gutman_cancer_2013}.

    Among the recent end-to-end systems proposed for histopathology image search, Yottixel \parencite{kalra_yottixel_2020}, SISH \parencite{chen_fast_2022}, RetCCL \parencite{wang_retccl_2023}, and HSHR \parencite{li_high-order_2023} have emerged as influential contenders, showcasing promising outcomes. Yottixel pioneered the processing of large-scale WSIs by introducing the concept of mosaics. Instead of extracting features from the entire WSI, Yottixel's approach involves extracting features from mosaic tiles using a DenseNet-based feature extractor. Additionally, Yottixel incorporates the notion of barcoding \parencite{tizhoosh_barcode_2015, tizhoosh_minmax_2016} to facilitate expedited retrieval by binarizing the extracted features. Similarly, SISH adopts a framework very similar to Yottixel but incorporates an additional VQ-VAE-based \parencite{oord_neural_2018} feature extractor. SISH also introduces advanced VEB tree-based \parencite{van_emde_boas_preserving_1977} indexing and ranking algorithms to enhance the quality of the retrieved samples. In contrast, RetCCL employs the mosaic concept as well, but uniquely converts WSIs to mosaics after extracting features from tiled WSIs. Moreover, RetCCL introduces an effective contrastive-based feature extractor to improve feature quality. Finally, HSHR expands the idea of using Self-Supervised Learning (SSL) to both extract the mosaics and creating hash codes from them. It uses SimCLR \parencite{chen_simple_2020} to train the feature extractor and uses MOCO \parencite{he_momentum_2020} to train the Cluster-Attention Hash Encoder (CaEncoder). The results are then processed to create a hypergraph which leads to similarity-based WSI retrieval. 

    The introduction of successive systems claiming to have achieved state-of-the-art performance in the search and retrieval of histopathology slides, often supported by statistical metrics demonstrating agreement with trained pathologists' judgments, raises the fundamental question of whether this problem has been satisfactorily addressed. Specifically, it prompts an inquiry into the readiness of these systems for deployment in clinical settings, where they can provide genuinely valuable information to pathologists, especially in challenging cases where even the most experienced group of pathologists struggle to reach a consensus. 

    In this case study, we evaluate these models on patient cases from our health system and several external datasets. Our objective is to provide a quantitative analysis of these models' performances on unseen slides while offering a qualitative assessment of the usefulness of these models in the clinical setting and potential areas of improvement. To ensure fairness, we constructed each model's database using a fixed number of slides from The Cancer Genome Atlas (TCGA) \parencite{weinstein_cancer_2013}, while employing the same feature extractors as published by the original authors.
    
    In subsequent sections, we provide an overview of our methods and approach towards implementation (\cref{sec:methods}). We then report our quantitative and and qualitative analysis of model performance (\cref{sec:results}). Finally, we discuss the current state of histopathology slide search engines and propose a set of minimal requirements for real-world deployment based on our findings (\cref{sec:discussion}).

    \section{Methods}
\label{sec:methods}

    \subsection{Search Engines}
    \label{subsec:searchengines}
        Search engines commonly comprise two fundamental components: indexing and database generation, as well as ranking and retrieval. Given the large-scale nature of the images involved in this study, feature extraction becomes imperative for effective indexing. In terms of ranking, a suitable similarity measure is crucial, followed by post-processing steps to ensure result quality. In the supplementary methods section, we provide an overview of the feature extraction techniques, database indexing approaches, employed similarity measures, and result ranking methodologies utilized by the four primary methods under investigation (\cref{fig:summary_of_methods}). Please be advised that all the hyper-parameters employed here are the parameters recommended by the authors of the models. We additionally share our code that we used to re implement methods that were not made available by the authors. 

        To summarize, The \textbf{Yottixel} method creates a mosaic of patches from whole slide images (WSIs), applies a feature extractor, KimiaNet \parencite{riasatian_fine-tuning_2021}, and generates binary codes, or barcodes, from the extracted features. These barcodes represent each WSI, form a database, and enable the retrieval of related slides or patches based on the median of the minimum Hamming distances \parencite{hamming_error_1950}.

The \textbf{SISH} method also generates a mosaic and uses DenseNet for feature extraction similar to Yottixel, but it further employs a pretrained VQ-VAE for index creation. Querying in SISH involves converting a slide into a mosaic, generating indices and features, and utilizing the "guided VEB search" algorithm to retrieve top slides based on Hamming distance.

\textbf{RetCCL} takes inspiration from both Yottixel and SISH but applies a unique approach by obtaining contrastive-based feature vectors for each patch within the segmented foreground tiles. The method employs a clustering-guided contrastive learning method with two InfoNCE losses to capture irregular regions in patches, which is particularly important given the prevalence of normal cells in WSIs \parencite{oord_representation_2019}.

Finally, \textbf{HSHR} first trains a encoder in a self-supervised manner on a small subset of patches extracted from database slides. By clustering the features from this encoder, it creates mosaics for each slide, and then passes the the features of mosaics to the CaEncoder. By following teh guidelines of MOCO \parencite{he_momentum_2020}, they train the desired CaEncoders and this way, they are able to create hashings and weights for each slides. Hashing and weights are then incorporated into building a hypergraph for the database. Every query slide is then considered a new node and hyperedge in this hypergraph database and similarity scores can be calculated for it. Per the authors' emphasize on global perception of WSIs, HSHR is not designed to be used with patch retrieval tasks.

A more detailed explanation of these models also can be found in Supplementary Materials (Supplementary Section \ref{sup:search_engine_methods}). Specifically, Supplementary Algorithms \ref{alg1} to \ref{alg4} would summarize the process a query slide would undergo in all the discussed methods. Moreover, time complexity of various stages of ranking and retrieval of these methods are also juxtaposed in \cref{tab:complexity_analysis}.

    \subsection{Database Slides}
        To ensure a fair comparison among all models, it was necessary to have consistent slides in the databases of each model. We constructed the database using slides available in TCGA \parencite{weinstein_cancer_2013}. Given our focus on lung, brain, and liver as primary sites for testing (see \cref{subsection:experiment}), it was essential to include slides from these sites in the databases. Additionally, to introduce a challenging aspect, slides from breast and colon were added to ensure that site retrieval experiments were not trivial. For each site, we randomly selected between 50 to 75 slides from subtypes containing at least 75 slides. The varying number of slides aimed to introduce class imbalance, mirroring real-world scenarios where some subtypes have more samples than others. Importantly, none of the slides in the database shared the same patient ID. The resulting database comprised 508 slides from 5 different sites and 8 different subtypes (\cref{tab:database}).

        It is worth noting that in each experiment, we utilized the pre-trained feature extractors provided by the respective authors (except for the backbone of HSHR, see Supplementary Section \ref{sup:search_engine_methods}). These feature extractors were trained on different datasets of varying sizes. The relatively small size of our database does not affect the performance of these models, as the only aspect influenced by data size is the feature extractor. As long as we have samples of the same class as the query within the database, a correctly functioning model should be capable of retrieving them.

        Due to preprocessing criteria, we were not able to include 6 slides for Yottixel, 1 slide for SISH,. 4 slides for HSHR in the database. These slides are listed in \cref{suptab:unprocessed_slides}.

    \subsection{Test Datasets}
        In order to conduct fair quantitative experiments and avoid data leakage, we needed to acquire test slides that were not seen by  the encoders of these models. \cref{tab:experiment_datasets} summarizes the all the datasets used in validation experiments. Except from the in house UCLA dataset, all other datasets are downloaded from the publicly available Cancer Imaging Archive database \parencite{clark_cancer_2013} (\cref{tab:test_data_links}). For all the datasets, we made sure not to have samples from the same patient using the patient identifiers provided.

        All UCLA slides were sourced from real clinical cases at our institution to best approximate real-world scenarios. Team members who were responsible for algorithmic implementation were blinded from the ground truth to reduce the likelihood of bias.

        \begin{table}[!ht]
            \centering
            \caption{Summary of test slides used for experiments. Abbreviations are based on \parencite{kalra_pan-cancer_2020}.}
            \label{tab:experiment_datasets}
            \begin{tabular}{p{3.5cm}p{3cm}p{4.5cm}p{1cm}p{1.5cm}}
                \toprule
                \textbf{Experiment} & \textbf{Slides} & \textbf{Dataset} & \textbf{Site} & \textbf{Diagnosis} \\
                \midrule
                \textbf{UCLA} & slide1 & In House & lung & LUAD \\
                & slide2 & In House & brain & LGG \\
                & slide3 & In House & liver & LIHC \\
                \midrule
                \textbf{Reader Study} & MSB-09151-01-11 & CMB-CRC & colon & COAD \\
                & MSB-09977-01-22 & CMB-LCA & lung & LUSC \\
                & Her2Pos\_Case\_66 & Yale Her2+ Cohort & breast & BRCA \\
                \midrule
                \textbf{Microscope Study} & 34 slides & CPTAC-GBM (Leica) & brain & GBM \\
                & 34 slides & UPENN-GBM (Hamamatsu) & brain & GBM \\
                \midrule
                \textbf{HER2+ prediction} & 93 slides & Yale Her2+ Cohort & breast & BRCA \\
                & 97 slides & Yale Her2- Cohort & breast & BRCA \\
                \midrule
                \textbf{Ablation} & 85 slides & Yale Trastuzumab Cohort & breast & BRCA \\
                \bottomrule
            \end{tabular}
        \end{table}

    \subsection{Experiments}
    \label{subsection:experiment}
        In general, we have three types of experiments: site (tissue) retrieval, subtype retrieval, and patch retrieval. We define “site” as the tissue of cancer origin and “subtype” as the final diagnosis, which is specific to the tissue type. For patch retrieval tasks, the models should return the closest patches to a query patch as opposed to WSIs to WSIs. For subtype and patch retrieval experimetns, we limited the search database to the slides with the same tissue type as the query.

        Consistent with prior work on WSI search algorithms, we chose majority top-k accuracy (mMV@K) and mean average precision (mAP@K) as our quantitative metric of performance, which returns the predicted label by majority vote amongst the top K slides. For tissue search, K has commonly been 10 for mMV and 5 for mAP. As with prior studies, we choose mMV@1,3,5 and mAP@3,5 for subtype search (See Supplementary Algorithms \ref{alg5} and \ref{alg6}).

        We have designed 5 experiments to validate the methods for different purposes. The UCLA experiment aims to bring qualitative evaluation to the 3 in house slides by evaluating them for all three tasks. Reader study is designed to bring pathologists' point of view to the quality retrieved patches by the models. The microscope study is intended to measure the robustness of the models' performances with respect to different microscope brands. The Her2+ prediction tries to answer the question whether there is evidence that the models would perform differently given different sub-subtype. And finally, since the authors' of Yottixel mentioned their model would perform better using KimiaNet instead of DenseNet pretrained on ImageNet, we designed the ablation study to measure this change in performance.

    \section{Results}
\label{sec:results}
            
\begin{table}[!ht]
    \centering
    \caption{Evaluation results of different methods on UCLA slides for primary site retrieval task.}
    \label{tab:primary_site_evaluation}
    \begin{tabular}{p{2.5cm}p{1.5cm}p{1.2cm}p{1.2cm}p{1.2cm}p{1.2cm}p{1.2cm}p{1.2cm}}
        \toprule
        \textbf{Method} & \textbf{UCLA Slides} & \textbf{MV@1} & \textbf{MV@3} & \textbf{MV@5} & \textbf{MV@10} & \textbf{AP@3} & \textbf{AP@5} \\
        \midrule
        \multirow{3}{*}{\shortstack{\textbf{YOTTIXEL} \\ \textbf{+ KimiaNet}}} & \textbf{Slide1} & 0 & 0 & 0 & 0 & 0 & 0 \\
        & \textbf{Slide2} & 1 & 1 & 1 & 1 & 1 & 1 \\
        & \textbf{Slide3} & 0 & 0 & 0 & 0 & 0 & 0 \\
        \midrule
        \multirow{3}{*}{\shortstack{\textbf{SISH +} \\ \textbf{DenseNet}}} & \textbf{Slide1} & 0 & - & - & - & 0 & 0 \\
        & \textbf{Slide2} & 1 & 1 & 1 & 1 & 1 & 0.888 \\
        & \textbf{Slide3} & 0 & 0 & - & - & 0 & 0 \\
        \midrule
        \multirow{3}{*}{\textbf{RetCCL}} & \textbf{Slide1} & 1 & 0 & 1 & 1 & 1 & 0.750 \\
        & \textbf{Slide2} & 0 & 1 & 1 & 1 & 0.583 & 0.679 \\
        & \textbf{Slide3} & 1 & 1 & 1 & 1 & 1 & 1 \\
        \midrule
        \multirow{3}{*}{\textbf{HSHR}} & \textbf{Slide1} & 1 & 1 & 1 & 1 & 1 & 1 \\
        & \textbf{Slide2} & 1 & 1 & 1 & 1 & 1 & 1 \\
        & \textbf{Slide3} & 0 & 0 & 0 & 0 & 0.500 & 0.500 \\
        \bottomrule
    \end{tabular}
\end{table}

\begin{table}[!t]
    \centering
    \caption{Evaluation results of different methods on UCLA slides for subtype retrieval task.}
    \label{tab:subtype_evaluation}
    \begin{tabular}{p{2.5cm}p{1.5cm}p{1cm}p{1cm}p{1cm}p{1cm}p{1cm}}
        \toprule
        \textbf{Method} & \textbf{UCLA Slides} & \textbf{MV@1} & \textbf{MV@3} & \textbf{MV@5} & \textbf{AP@3} & \textbf{AP@5} \\
        \midrule
        \multirow{3}{*}{\shortstack{\textbf{YOTTIXEL} \\ \textbf{+ KimiaNet}}} & \textbf{Slide1} & 0 & 0 & 1 & 0.500 & 0.533 \\
        & \textbf{Slide2} & 0 & 0 & 0 & 0 & 0 \\
        & \textbf{Slide3} & 1 & 1 & 1 & 1 & 1 \\
        \midrule
        \multirow{3}{*}{\shortstack{\textbf{SISH +} \\ \textbf{DenseNet}}} & \textbf{Slide1} & 1 & - & - & 1 & 1 \\
        & \textbf{Slide2} & 0 & 0 & 0 & 0 & 0 \\
        & \textbf{Slide3} & 1 & 1 & 1 & 1 & 1 \\
        \midrule
        \multirow{3}{*}{\textbf{RetCCL}} & \textbf{Slide1} & 1 & 1 & 1 & 1 & 1 \\
        & \textbf{Slide2} & 1 & 1 & 1 & 1 & 1 \\
        & \textbf{Slide3} & 1 & 0 & 0 & 1 & 0.700 \\
        \midrule
        \multirow{3}{*}{\textbf{HSHR}} & \textbf{Slide1} & 1 & 1 & 1 & 1 & 0.867 \\
        & \textbf{Slide2} & 1 & 0 & 0 & 1 & 1 \\
        & \textbf{Slide3} & 0 & 0 & 0 & 0.500 & 0.500 \\
        \bottomrule
    \end{tabular}
\end{table}

\begin{figure}[!ht]
\label{fig:testt}
    \centering
    \includegraphics[scale=0.15]{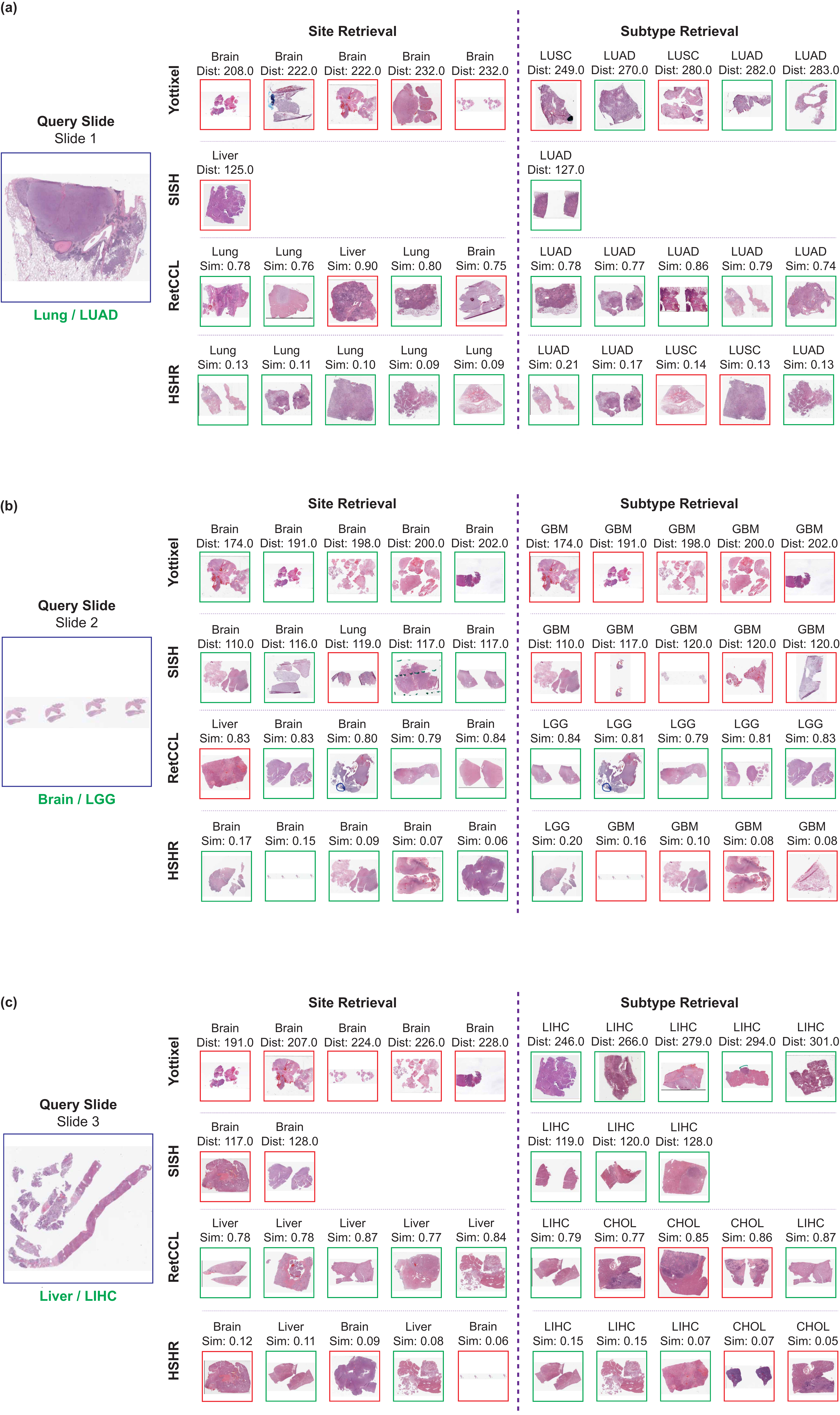}
    \caption{Results of site retrieval (left) and sub-type retrieval (right) at slide level for all three test slides. Correct labels are printed in green under query slides. Green border means correct label; red border means wrong label. For details about distances and similarities, see \cref{subsec:searchengines}. }
\end{figure}

Detailed results are included in the supplement, and a concise summary is provided here.  

    \subsection{Tissue and subtype retrieval}

A quantitative analysis of model performance on our patient cases is reported in \cref{tab:primary_site_evaluation} and \cref{tab:subtype_evaluation}. While we cannot arrive at definitive conclusions on algorithmic performance due to the few number of cases here, we observed higher mMV and mAP metrics for RetCCL on both tissue and subtype retrieval relative to the other models.  

\cref{suptab:primary_site_evaluation_rest} and \cref{suptab:primary_subtype_evaluation_rest} show the results of our quantitative evaluations on GBM and BRCA datasets encompassing 344 unseen slides. Of note, UPENN GBM was not feasible for RetCCL due to extensive computational burdens. When testing the models on brain tissue retrieval, we found that RetCCL had the highest performance at mMV @ 10 although there is almost a 15 point drop in performance versus the mMV @ 10 score reported by the authors at 90.21. For subtyping on GBM versus LGG, we see that HSHR and SISH are the top performers at mMV@5 for GBM versus LBB subtyping, which are lower than prior work by the authors of HSHR showing a mMV@5 of 0.937 and 0.916 for HSHR and SISH, respectively, when tested on 3580 TGCA GBM and LGG slides. 

For breast tissue retrieval on the Yale Trastuzumab dataset, Yotixxel performed the best on both metrics with a mMV@10 of 0.588 and mAP@5 of 0.650. Of note, the authors of RetCCL previously showed that Yotixxel achieved a mMV@10 of 0.663 relative to RetCCL's score of 0.914 on a set of frozen WSIs. Subtype search on the BRCA dataset was not performed as all of the cases were of the same diagnosis. 

\subsection{Visual review of query results}
To better investigate discrepancies in performance, we reviewed the top five ranked results on subtyping and tissue search for three different WSI slides from our own patient cases as illustrated in Figure 2. Additional details on patient cases and slide preparation methods can be found in the Supplementary Section on Patient Cases. 

In Figure 2, we see several errors made on tissue and subtype search. We also review of patch-level results for Yotixxel, SISH, and RetCCL on two patches from our LUAD case, one showcasing tumoral tissue and the other with normal alveolar tissue (Figure 3). HSHR was excluded due to its inability for patch-level search. We found that all three algorithms are capable of retrieving patches containing tumoral and alveolar tissue but there were visual discrepancies in some of more granular features on slides. For example, while all models retrieve patches corresponding to alveoli, all three models return patches with varying degrees of necrosis, inflammation, and hyperplasia, which may point pathologists towards different diagnoses and treatments.  

\subsection{Reader study}
We next compared model performance on patch-level retrieval results qualitatively. Supplementary Figure 1 shows the Mean Opinion Score (MOS) of seven pathologists on the top three ranked results when querying a patch containing tumor from three WSI H\&E slides. We include Yottixel, RetCCL, and SISH but not HSHR given the latter’s inability to perform patch-level search. For consistency, we used MOS as an evaluation metric based on prior studies on the quality of WSI search results. We found SISH had overlapping performance with Yotixxel and RetCCL due to higher variance but when comparing RetCCL versus Yotixxel, we see a statistically significant improvement in RetCCL performance. However, the Fleiss' Kappa for all algorithms combined was 0.131 suggesting low rates of agreement amongst pathologists. Quantitative performance metrics are provided in Supplementary Tables 5 and 6 on our reader study slides. 

\subsection{HER2+ prediction}
To test the richness of feature representations, we compared the ability of the four models in distinguishing between HER2+ and HER2- BRCA (Supplementary Figures 3,4,5, \& 6). At present, immunohistochemistry is required for the interpretation of HER2 status although recent work using a CNN-based architecture was capable of predicting HER2 positivity with an AUC of 0.81 on untested datasets (\cite{farahmand2022deep}). All models tended to predict HER2- slides with greater precision despite the nearly 50-50 distribution of HER2- and HER2+ cases in the dataset, suggesting an ability to learn subtle features specific to HER2 status.

For this experiment, we calculated MV@10 and AP@5 for tissue retrieval task in both Her2+ and Her2- cohorts. Then using the Shapiro-Wilk normality test, and Levene's homogeneity of variances test, we checked the distribution of these two metrics. Since they did not pass the Independent T test requirements, we conducted a non-parametric Mann-Whitney U Test. We concluded that all 4 models 
would have a better performance in terms of AP@5 on Her2- cohort for the site retrieval task. However, for MV@10, the evidence was only significant for Yottixel and HSHR (\cref{tab:u_test_results}).

\subsection{Ablation study}

Figure \ref{fig:algorithm_comparison} shows comparative analysis of Yottixel and SISH algorithms using two distinct networks, Kima Net and Densenet, on the Yale Trastuzumab dataset is presented. The figure is subdivided into two subplots, each illustrating the performance of a base algorithm, Yottixel or SISH, evaluated across six different metrics: mmV @ 1, 3, 5, 10 and mAP @ 3 and 5. Configurations employing Kima Net generally outperform those using Densenet across our evaluation metrics. Disparities in performance underscore the impact of the network choice on the efficacy of the algorithms, highlighting the importance of optimal network-algorithm pairing.

\subsection{Microscope study} 
We investigated differences in performance for Yotixxel, SISH, and HSHR on two GBM datasets utilizing different microscopes, UPENN and CPTAC GBM, to test model generalizability. As RetCCL was not compatible with the UPENN dataset, it was excluded from this experiment. We found that all three algorithms had overlapping results with p-values of $>$ 0.05 across both subtype and tissue retrieval at mMV @ 5 and 10 on both GBM datasets, indicating that differences in microscopes did not affect performance. Full results can be found in the \cref{tab:u_test_results_microscope}.

        \begin{figure}[!ht]
            \centering
            \includegraphics[width=\textwidth]{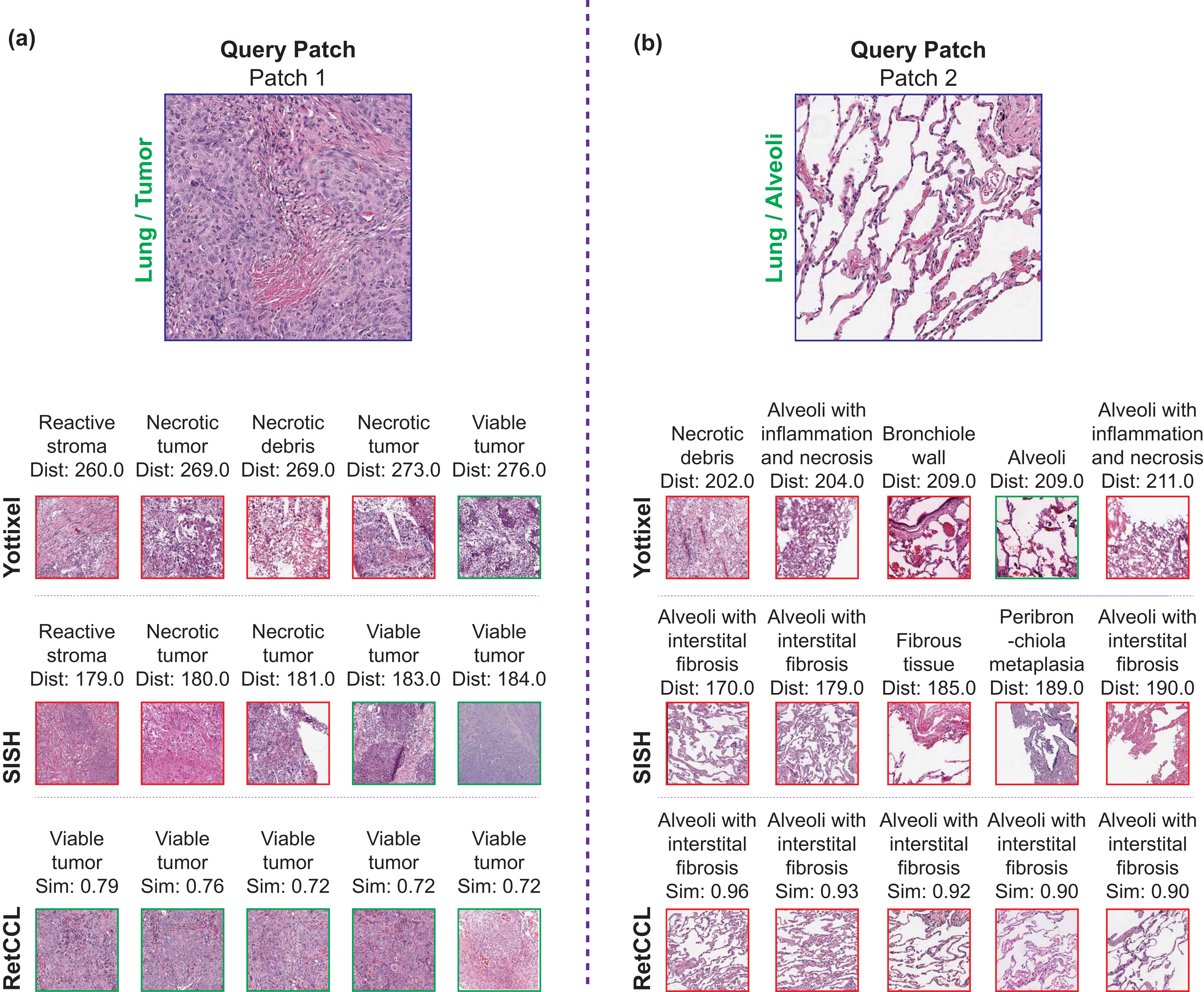}
            \caption{Results of patch retrieval for two patches from Slide 1. Correct labels are printed in green to the left of query patches. Green border means correct label; red border means wrong label. For details about distances and similarities, see \cref{subsec:searchengines}.}
            \label{fig:test_patch}
        \end{figure}

    \section{Discussion}
\label{sec:discussion}

    In this case study, we evaluated the performance and clinical utility of state-of-the-art histopathology slide search engines, Yotixxel, SISH, and RetCCL and HSHR. To our knowledge, this is the first independent, external validation of these four models. While all models demonstrate significant advancements in the field, we also noticed clinically significant shortcomings.
    
    Based on our findings, we propose a framework of minimal requirements to facilitate the development of these systems both fairly and transparently while minimizing patient harm and maximizing clinical utility:

      \begin{enumerate}

\item \textbf{Richness of feature representations:} 
We observed several clinically relevant inconsistencies across our queried and retrieved patches, suggesting the need for improvements in feature extraction. Our ablation study also highlights the importance of encoder architecture on model performance. Large pre-trained models such as Virchow (\cite{vorontsov2023virchow}), with extensive training on 1.5 million images, are a potential solution, similar to natural image processing models such as VGG16 (\cite{simonyan2015deep}) and ResNet-50 (\cite{he2015deep}). This strategy can also facilitate the prioritization of downstream tasks while alleviating extensive computational burdens in the pre-training stage. 

     \item \textbf{Systematic and rigorous evaluations:} 
We found it challenging to evaluate relative model performance due to variable results across the four algorithms in our experiments. For example, RetCCL had the highest performance on brain tissue retrieval but Yottixel had better performance on breast tissue. Likewise, while we found that pathologists tended to rank patches retrieved by RetCCL highly, due to large inter-observer variation on quality, relative performance was still uncertain. Our findings highlight the need for systematic and rigorous methodologies for model evaluation.  
    
    \item \textbf{Robustness:} A vital measure of clinical applicability is model performance across diverse clinical populations and environments, as in the case of real-world health systems. This is motivated by our findings that all models experienced a notable decrease in performance when tested against unseen datasets. However, we were pleased to find that batch level effects from external factors such as microscope type did not result in statistically significant differences on tissue and subtyping performance. We propose that all future models be validated for their generalizability in addition to precision. 

    \item \textbf{Replicability:} We found variable degrees of replicability across the models evaluated in this work. Some were constructed for specific slide ratios and resolutions, while others were capable of handling variable slide formats. Models also differed in terms of ease of applicability and transparency due to differences in the availability of model development code and indexed databases. Model transparency fosters scientific integrity and accelerates the pace of innovation through collective efforts and unbiased evaluations.

    \item \textbf{Clinical benefit:} For clinical adoption, the unique aspects of each model and their value need to be clear for end-users. To date, the ongoing development of WSI search algorithms has focused on achieving stepwise improvements in performance on tissue and subtyping search. While these are worthy goals, we propose the inclusion of new tasks based on existing challenges in the field of pathology, such as helping to determine the tissue of cancerous origin in the metastatic setting.
    
    \item \textbf{Computational Efficiency:} Clinical applicability not only requires theoretical efficiency, but also hinges on the performance of these systems in a real-world, high-demand environment. Thus, discussions on computational efficiency should reflect the realities of clinical implementation. Especially, we need to make sure the processes of querying an expanding the database are really efficient. These two processes are of utmost importance for a sustainable search engine. We recommend the database indexing algorithms to be online (i.e. adding a new data point does not require compiling the whole database from scratch). Also, one shortcoming of models like RetCCL, SISH, and Yottixel is that they generate mosaics as a percentage of number of patches in a cluster. Now, if like GBM UPENN, slides become very large, number of patches becomes unnecessarily large and query time becomes very long. We recommend models such as HSHR at use a fixed size for patches for this matter. 

\end{enumerate}

  As the field of digital pathology continues to evolve, we anticipate exciting developments in the near future. These will likely include more efficient and reliable systems for indexing and searching of histopathology slides, increasingly robust algorithms for feature extraction, and potentially transformative diagnostic tools. Given the high-stakes nature of patient care, we anticipate a significant amount of work ahead to ensure the validity of these models prior to clinical adoption. As we continue to make strides in the development of histopathology slide search engines, our proposed criteria ensures that these systems are not only theoretically sound but also ready for meaningful clinical adoption.

    \section*{Code and Data Availability}
    The test slides along with the updated source codes for all three methods used to generate the results can be found at \href{https://github.com/jacobluber/PathologySearchComparison}{github.com/jacobluber/PathologySearchComparison}. All other data used in databases are publicly available at \href{https://portal.gdc.cancer.gov/}{portal.gdc.cancer.gov} and \href{https://www.cancerimagingarchive.net}{www.cancerimagingarchive.net/}. The list of data included in the database can also be found in the github repository above.
    \section*{Funding}
This work was supported by a University of Texas System Rising STARs Award (J.M.L) and the CPRIT First
Time Faculty Award (J.M.L).

\section*{Competing Interests}
No competing interests are disclosed by the authors.

\section*{Acknowledgements}
  We thank the following for their contributions: David Fernandez-Hazoury, M.D., Department of Pathology, Harbor-UCLA Medical Center; Kenechukwu Ojukwu, M.D., M.P.P.,Bone/Soft Tissue Fellow, Department of Pathology and Laboratory Medicine, National Clinician Scholars Program (NCSP) Fellow; Alex Oliveira-Kowaleski, M.D., UCLA Department of Pathology and Laboratory Medicine; Julie Y. Kim, D.O., M.S., Harbor-UCLA Medical Center; Daniel P. Stefanko, M.D. Ph.D., UCLA Department of Pathology and Laboratory Medicine.

    \printbibliography

@article{farahmand2022deep,
  title={Deep learning trained on hematoxylin and eosin tumor region of Interest predicts HER2 status and trastuzumab treatment response in HER2+ breast cancer},
  author={Farahmand, Saman and Fernandez, Aileen I and Ahmed, Fahad Shabbir and Rimm, David L and Chuang, Jeffrey H and Reisenbichler, Emily and Zarringhalam, Kourosh},
  journal={Modern Pathology},
  volume={35},
  number={1},
  pages={44--51},
  year={2022},
  publisher={Nature Publishing Group US New York}
}

@misc{he2015deep,
      title={Deep Residual Learning for Image Recognition}, 
      author={Kaiming He and Xiangyu Zhang and Shaoqing Ren and Jian Sun},
      year={2015},
      eprint={1512.03385},
      archivePrefix={arXiv},
      primaryClass={cs.CV}
}

@misc{simonyan2015deep,
      title={Very Deep Convolutional Networks for Large-Scale Image Recognition}, 
      author={Karen Simonyan and Andrew Zisserman},
      year={2015},
      eprint={1409.1556},
      archivePrefix={arXiv},
      primaryClass={cs.CV}
}

@misc{vorontsov2023virchow,
      title={Virchow: A Million-Slide Digital Pathology Foundation Model}, 
      author={Eugene Vorontsov and Alican Bozkurt and Adam Casson and George Shaikovski and Michal Zelechowski and Siqi Liu and Philippe Mathieu and Alexander van Eck and Donghun Lee and Julian Viret and Eric Robert and Yi Kan Wang and Jeremy D. Kunz and Matthew C. H. Lee and Jan Bernhard and Ran A. Godrich and Gerard Oakley and Ewan Millar and Matthew Hanna and Juan Retamero and William A. Moye and Razik Yousfi and Christopher Kanan and David Klimstra and Brandon Rothrock and Thomas J. Fuchs},
      year={2023},
      eprint={2309.07778},
      archivePrefix={arXiv},
      primaryClass={eess.IV}
}

@misc{oord_representation_2019,
	title = {Representation {Learning} with {Contrastive} {Predictive} {Coding}},
	url = {http://arxiv.org/abs/1807.03748},
	doi = {10.48550/arXiv.1807.03748},
	urldate = {2023-06-27},
	publisher = {arXiv},
	author = {Oord, Aaron van den and Li, Yazhe and Vinyals, Oriol},
	month = jan,
	year = {2019},
	note = {arXiv:1807.03748 [cs, stat]},
}

@article{hamming_error_1950,
  title={Error detecting and error correcting codes},
  author={Hamming, Richard W},
  journal={The Bell system technical journal},
  volume={29},
  number={2},
  pages={147--160},
  year={1950},
  publisher={Nokia Bell Labs}
}

@article{weinstein_cancer_2013,
	title = {The {Cancer} {Genome} {Atlas} {Pan}-{Cancer} analysis project},
	volume = {45},
	copyright = {2013 The Author(s)},
	issn = {1546-1718},
	url = {https://www.nature.com/articles/ng.2764},
	doi = {10.1038/ng.2764},
	language = {en},
	number = {10},
	urldate = {2023-06-23},
	journal = {Nature Genetics},
	author = {Weinstein, John N. and Collisson, Eric A. and Mills, Gordon B. and Shaw, Kenna R. Mills and Ozenberger, Brad A. and Ellrott, Kyle and Shmulevich, Ilya and Sander, Chris and Stuart, Joshua M.},
	month = oct,
	year = {2013},
	note = {Number: 10
Publisher: Nature Publishing Group},
	pages = {1113--1120},
}

@article{van_emde_boas_preserving_1977,
	title = {Preserving order in a forest in less than logarithmic time and linear space},
	volume = {6},
	issn = {0020-0190},
	url = {https://www.sciencedirect.com/science/article/pii/002001907790031X},
	doi = {10.1016/0020-0190(77)90031-X},
	language = {en},
	number = {3},
	urldate = {2023-06-23},
	journal = {Information Processing Letters},
	author = {van Emde Boas, P.},
	month = jun,
	year = {1977},
	pages = {80--82},
}

@article{gutman_cancer_2013,
	title = {Cancer {Digital} {Slide} {Archive}: an informatics resource to support integrated in silico analysis of {TCGA} pathology data},
	volume = {20},
	issn = {1527-974X},
	shorttitle = {Cancer {Digital} {Slide} {Archive}},
	doi = {10.1136/amiajnl-2012-001469},
	language = {eng},
	number = {6},
	journal = {Journal of the American Medical Informatics Association: JAMIA},
	author = {Gutman, David A. and Cobb, Jake and Somanna, Dhananjaya and Park, Yuna and Wang, Fusheng and Kurc, Tahsin and Saltz, Joel H. and Brat, Daniel J. and Cooper, Lee A. D.},
	year = {2013},
	pmid = {23893318},
	pmcid = {PMC3822112},
	pages = {1091--1098},
}

@article{kalra_yottixel_2020,
	title = {Yottixel - {An} {Image} {Search} {Engine} for {Large} {Archives} of {Histopathology} {Whole} {Slide} {Images}},
	volume = {65},
	issn = {1361-8423},
	doi = {10.1016/j.media.2020.101757},
	language = {eng},
	journal = {Medical Image Analysis},
	author = {Kalra, Shivam and Tizhoosh, H. R. and Choi, Charles and Shah, Sultaan and Diamandis, Phedias and Campbell, Clinton J. V. and Pantanowitz, Liron},
	month = oct,
	year = {2020},
	pmid = {32623275},
	pages = {101757},
}

@article{wang_retccl_2023,
	title = {{RetCCL}: {Clustering}-guided contrastive learning for whole-slide image retrieval},
	volume = {83},
	issn = {1361-8415},
	shorttitle = {{RetCCL}},
	url = {https://www.sciencedirect.com/science/article/pii/S1361841522002730},
	doi = {10.1016/j.media.2022.102645},
	language = {en},
	urldate = {2023-05-11},
	journal = {Medical Image Analysis},
	author = {Wang, Xiyue and Du, Yuexi and Yang, Sen and Zhang, Jun and Wang, Minghui and Zhang, Jing and Yang, Wei and Huang, Junzhou and Han, Xiao},
	month = jan,
	year = {2023},
	pages = {102645},
}

@article{chen_fast_2022,
	title = {Fast and scalable search of whole-slide images via self-supervised deep learning},
	volume = {6},
	copyright = {2022 The Author(s)},
	issn = {2157-846X},
	url = {https://www.nature.com/articles/s41551-022-00929-8},
	doi = {10.1038/s41551-022-00929-8},
	language = {en},
	number = {12},
	urldate = {2023-05-11},
	journal = {Nature Biomedical Engineering},
	author = {Chen, Chengkuan and Lu, Ming Y. and Williamson, Drew F. K. and Chen, Tiffany Y. and Schaumberg, Andrew J. and Mahmood, Faisal},
	month = dec,
	year = {2022},
	note = {Number: 12
Publisher: Nature Publishing Group},
	pages = {1420--1434},
}

@article{kalra_pan-cancer_2020,
	title = {Pan-cancer diagnostic consensus through searching archival histopathology images using artificial intelligence},
	volume = {3},
	issn = {2398-6352},
	url = {https://www.ncbi.nlm.nih.gov/pmc/articles/PMC7064517/},
	doi = {10.1038/s41746-020-0238-2},
	urldate = {2023-05-11},
	journal = {NPJ Digital Medicine},
	author = {Kalra, Shivam and Tizhoosh, H. R. and Shah, Sultaan and Choi, Charles and Damaskinos, Savvas and Safarpoor, Amir and Shafiei, Sobhan and Babaie, Morteza and Diamandis, Phedias and Campbell, Clinton J. V. and Pantanowitz, Liron},
	month = mar,
	year = {2020},
	pmid = {32195366},
	pmcid = {PMC7064517},
	pages = {31},
}

@misc{tizhoosh_minmax_2016,
	title = {{MinMax} {Radon} {Barcodes} for {Medical} {Image} {Retrieval}},
	url = {http://arxiv.org/abs/1610.00318},
	doi = {10.48550/arXiv.1610.00318},
	urldate = {2023-05-11},
	publisher = {arXiv},
	author = {Tizhoosh, H. R. and Zhu, Shujin and Lo, Hanson and Chaudhari, Varun and Mehdi, Tahmid},
	month = oct,
	year = {2016},
	note = {arXiv:1610.00318 [cs]},
}

@inproceedings{tizhoosh_barcode_2015,
	title = {Barcode annotations for medical image retrieval: {A} preliminary investigation},
	shorttitle = {Barcode annotations for medical image retrieval},
	doi = {10.1109/ICIP.2015.7350913},
	booktitle = {2015 {IEEE} {International} {Conference} on {Image} {Processing} ({ICIP})},
	author = {Tizhoosh, H. R.},
	month = sep,
	year = {2015},
	pages = {818--822},
}

@misc{riasatian_fine-tuning_2021,
	title = {Fine-{Tuning} and {Training} of {DenseNet} for {Histopathology} {Image} {Representation} {Using} {TCGA} {Diagnostic} {Slides}},
	url = {http://arxiv.org/abs/2101.07903},
	doi = {10.48550/arXiv.2101.07903},
	urldate = {2023-05-23},
	publisher = {arXiv},
	author = {Riasatian, Abtin and Babaie, Morteza and Maleki, Danial and Kalra, Shivam and Valipour, Mojtaba and Hemati, Sobhan and Zaveri, Manit and Safarpoor, Amir and Shafiei, Sobhan and Afshari, Mehdi and Rasoolijaberi, Maral and Sikaroudi, Milad and Adnan, Mohd and Shah, Sultaan and Choi, Charles and Damaskinos, Savvas and Campbell, Clinton JV and Diamandis, Phedias and Pantanowitz, Liron and Kashani, Hany and Ghodsi, Ali and Tizhoosh, H. R.},
	month = jan,
	year = {2021},
	note = {arXiv:2101.07903 [eess]},
}

@misc{oord_neural_2018,
	title = {Neural {Discrete} {Representation} {Learning}},
	url = {http://arxiv.org/abs/1711.00937},
	doi = {10.48550/arXiv.1711.00937},
	urldate = {2023-06-07},
	publisher = {arXiv},
	author = {Oord, Aaron van den and Vinyals, Oriol and Kavukcuoglu, Koray},
	month = may,
	year = {2018},
	note = {arXiv:1711.00937 [cs]},
}

@article{hegde_similar_2019,
	title = {Similar image search for histopathology: {SMILY}},
	volume = {2},
	copyright = {2019 The Author(s)},
	issn = {2398-6352},
	shorttitle = {Similar image search for histopathology},
	url = {https://www.nature.com/articles/s41746-019-0131-z},
	doi = {10.1038/s41746-019-0131-z},
	language = {en},
	number = {1},
	urldate = {2023-06-22},
	journal = {npj Digital Medicine},
	author = {Hegde, Narayan and Hipp, Jason D. and Liu, Yun and Emmert-Buck, Michael and Reif, Emily and Smilkov, Daniel and Terry, Michael and Cai, Carrie J. and Amin, Mahul B. and Mermel, Craig H. and Nelson, Phil Q. and Peng, Lily H. and Corrado, Greg S. and Stumpe, Martin C.},
	month = jun,
	year = {2019},
	note = {Number: 1
Publisher: Nature Publishing Group},
	pages = {1--9},
}

@article{li_large-scale_2018,
	title = {Large-scale retrieval for medical image analytics: {A} comprehensive review},
	volume = {43},
	issn = {1361-8415},
	shorttitle = {Large-scale retrieval for medical image analytics},
	url = {https://www.sciencedirect.com/science/article/pii/S136184151730138X},
	doi = {10.1016/j.media.2017.09.007},
	language = {en},
	urldate = {2023-06-22},
	journal = {Medical Image Analysis},
	author = {Li, Zhongyu and Zhang, Xiaofan and Müller, Henning and Zhang, Shaoting},
	month = jan,
	year = {2018},
	pages = {66--84},
}

@article{lew_content-based_2006,
	title = {Content-based multimedia information retrieval: {State} of the art and challenges},
	volume = {2},
	issn = {1551-6857},
	shorttitle = {Content-based multimedia information retrieval},
	url = {https://dl.acm.org/doi/10.1145/1126004.1126005},
	doi = {10.1145/1126004.1126005},
	number = {1},
	urldate = {2023-06-23},
	journal = {ACM Transactions on Multimedia Computing, Communications, and Applications},
	author = {Lew, Michael S. and Sebe, Nicu and Djeraba, Chabane and Jain, Ramesh},
	month = feb,
	year = {2006},
	pages = {1--19},
}

@article{li_high-order_2023,
	title = {High-Order Correlation-Guided Slide-Level Histology Retrieval With Self-Supervised Hashing},
	volume = {45},
	issn = {1939-3539},
	doi = {10.1109/TPAMI.2023.3269810},
	pages = {11008--11023},
	number = {9},
	journaltitle = {{IEEE} Transactions on Pattern Analysis and Machine Intelligence},
	author = {Li, Shengrui and Zhao, Yining and Zhang, Jun and Yu, Ting and Zhang, Ji and Gao, Yue},
	date = {2023-09},
	note = {Conference Name: {IEEE} Transactions on Pattern Analysis and Machine Intelligence},
}

@misc{chen_simple_2020,
	title = {A Simple Framework for Contrastive Learning of Visual Representations},
	url = {http://arxiv.org/abs/2002.05709},
	doi = {10.48550/arXiv.2002.05709},
	number = {{arXiv}:2002.05709},
	publisher = {{arXiv}},
	author = {Chen, Ting and Kornblith, Simon and Norouzi, Mohammad and Hinton, Geoffrey},
	urldate = {2023-09-25},
	date = {2020-06-30},
	eprinttype = {arxiv},
	eprint = {2002.05709 [cs, stat]},
}

@misc{he_momentum_2020,
	title = {Momentum Contrast for Unsupervised Visual Representation Learning},
	url = {http://arxiv.org/abs/1911.05722},
	doi = {10.48550/arXiv.1911.05722},
	number = {{arXiv}:1911.05722},
	publisher = {{arXiv}},
	author = {He, Kaiming and Fan, Haoqi and Wu, Yuxin and Xie, Saining and Girshick, Ross},
	urldate = {2023-09-25},
	date = {2020-03-23},
	eprinttype = {arxiv},
	eprint = {1911.05722 [cs]},
}

@misc{he_deep_2015,
	title = {Deep Residual Learning for Image Recognition},
	url = {http://arxiv.org/abs/1512.03385},
	doi = {10.48550/arXiv.1512.03385},
	number = {{arXiv}:1512.03385},
	publisher = {{arXiv}},
	author = {He, Kaiming and Zhang, Xiangyu and Ren, Shaoqing and Sun, Jian},
	urldate = {2023-09-25},
	date = {2015-12-10},
	eprinttype = {arxiv},
	eprint = {1512.03385 [cs]},
}

@article{clark_cancer_2013,
	title = {The Cancer Imaging Archive ({TCIA}): Maintaining and Operating a Public Information Repository},
	volume = {26},
	issn = {1618-727X},
	url = {https://doi.org/10.1007/s10278-013-9622-7},
	doi = {10.1007/s10278-013-9622-7},
	shorttitle = {The Cancer Imaging Archive ({TCIA})},
	pages = {1045--1057},
	number = {6},
	journaltitle = {Journal of Digital Imaging},
	shortjournal = {J Digit Imaging},
	author = {Clark, Kenneth and Vendt, Bruce and Smith, Kirk and Freymann, John and Kirby, Justin and Koppel, Paul and Moore, Stephen and Phillips, Stanley and Maffitt, David and Pringle, Michael and Tarbox, Lawrence and Prior, Fred},
	urldate = {2023-09-26},
	date = {2013-12-01},
	langid = {english},
}

    \newpage
    \appendix
    \setcounter{sfigure}{0}
    \setcounter{stable}{0}
    \setcounter{sequation}{0}
    \section*{Supplementary Material}

\subsection*{Datasets}

\begin{stable}[!h]
    \centering
    \caption{Summary of database used for comparison experiments. Abbreviations are based on \parencite{kalra_pan-cancer_2020}.}
    \label{tab:database}
    \begin{threeparttable}
        \begin{tabular}{p{2cm}p{5cm}p{2cm}p{2cm}p{1.8cm}}
            \toprule
            \textbf{Primary Site} & \textbf{Project Name (Subtype)} & \textbf{Abbr.} & \textbf{Num. Slides} & \textbf{Num. Selected Slides} \\
            \midrule
            Brain & Glioblastoma Multiforme & GBM & 2040 & 61 \\ 
            & Brain Lower Grade Glioma & LGG & 1543 & 69 \\
            & Lymphoid Neasm Diffuse Large B-cell Lymphoma & DLBC & 4 & 0 \\
            \midrule
            Breast & Breast Invasive Carcinoma & BRCA & 2704 & 72 \\
            & Lymphoid Neasm Diffuse Large B-cell Lymphoma & DLBC & 2 & 0 \\
            \midrule
            Bronchus and lung & Lung Adenocarcinoma & LUAD & 1359 & 68 \\
            & Lung Squamous Cell Carcinoma & LUSC & 1265 & 57 \\
            & Mesothelioma & MESO & 2 & 0 \\
            \midrule
            Colon & Colon Adenocarcinoma/Rectum Adenocarcinoma\tnote{a} & COAD/READ & 1307+18 & 59 (COAD) \\
            & Lymphoid Neasm Diffuse Large B-cell Lymphoma & DLBC & 6 & 0 \\
            & Sarcoma & SARC & 4 & 0 \\
            \midrule
            Liver and intrahepatic bile ducts & Liver Hepatocellular Carcinoma & LIHC & 778 & 72 \\
            & Cholangiocarcinoma & CHOL & 80 & 50 \\
            \bottomrule
        \end{tabular}
        \begin{tablenotes}
            \item[a] Although from pathologist point of view Colon Adenocarcinoma and Rectum Adenocarcinoma are genetically and morphologically the same entity, TCGA considers them different projects.
        \end{tablenotes}
    \end{threeparttable}
\end{stable}

\begin{stable}[!h]
    \centering
    \caption{Access links to the test datasets used in experiments.}
    \label{tab:test_data_links}
        \begin{tabular}{p{3cm}p{12cm}}
            \toprule
            \textbf{Dataset} & \textbf{link} \\
            \midrule
            CMB-CRC  & \href{https://wiki.cancerimagingarchive.net/pages/viewpage.action?pageId=93257955}{wiki.cancerimagingarchive.net/pages/viewpage.action?pageId=93257955} \\
            CMB-LCA & \href{https://wiki.cancerimagingarchive.net/pages/viewpage.action?pageId=93258420}{wiki.cancerimagingarchive.net/pages/viewpage.action?pageId=93258420} \\
             Yale Her2 & \href{https://wiki.cancerimagingarchive.net/pages/viewpage.action?pageId=119702524}{wiki.cancerimagingarchive.net/pages/viewpage.action?pageId=119702524} \\
             Yale Trastuzumab & \href{https://wiki.cancerimagingarchive.net/pages/viewpage.action?pageId=119702524}{wiki.cancerimagingarchive.net/pages/viewpage.action?pageId=119702524} \\
             CPTAC-GBM  & \href{https://wiki.cancerimagingarchive.net/pages/viewpage.action?pageId=30671232}{wiki.cancerimagingarchive.net/pages/viewpage.action?pageId=30671232} \\
             UPENN-GBM & \href{https://wiki.cancerimagingarchive.net/pages/viewpage.action?pageId=70225642}{wiki.cancerimagingarchive.net/pages/viewpage.action?pageId=70225642}\\
            \bottomrule
        \end{tabular}
\end{stable}

\begin{stable}[!h]
            \centering
            \caption{Unprocessed slides in the database for each model.}
            \label{suptab:unprocessed_slides}
            \begin{tabular}{p{2cm}p{7cm}}
                \toprule
                \textbf{Model} & \textbf{Slide IDs} \\
                \midrule
                \multirow{6}{*}{\textbf{Yottixel}} & 9ecf91d4-0d9e-4400-bf38-99420acd14cc \\
                & f18b6fc0-6f40-4f0d-82ef-0b092a21b6bf \\
                & 846087b8-f70c-4970-a1b7-24d403229801 \\
                & c95681f3-53d4-4b15-833d-ff68f171965e \\
                & 71dc7ba0-a623-4aaf-9502-f2fe9d188401 \\
                & 2dc5d0b4-04ff-4731-bda1-8ad7cd0fa345 \\
                \midrule
                \textbf{SISH} & 2dc5d0b4-04ff-4731-bda1-8ad7cd0fa345 \\
                \midrule
                \textbf{RetCCL} & All slides in the database processed. \\
                \midrule
                \multirow{4}{*}{\textbf{HSHR}} & f2d5aa37-d9ce-4264-a447-fc69dd0d7d85 \\
                & a2658e39-e476-44b2-99ee-118056cf6201 \\
                & f84130fe-4853-4252-a292-9372aeea4a5d \\
                & 22904f9d-0788-463c-9961-02629cf9a85f \\
                \bottomrule
            \end{tabular}
        \end{stable}
        
 \subsection*{H\&E Staining and Preparation}
 
        Tissues were stained with Harris’ hematoxylin solution for \SI{6}{\hour} at a temperature of \SIrange{60}{70}{\celsius} and were then rinsed in tap water until the water was colorless. Next, 10\% acetic acid and 85\% ethanol in water were used to differentiate the tissue 2 times for \SI{2}{\hour} and \SI{10}{\hour}, and the tissues were rinsed with tap water. In the bluing step, we soaked the tissue in saturated lithium carbonate solution for \SI{12}{\hour} and then rinsed it with tap water. Finally, staining was performed with eosin Y ethanol solution for \SI{48}{\hour}. Tissues were dehydrated with 95\% ethanol twice for \SI{0.5}{\hour}, and then soaked in xylene for \SI{1}{\hour} at \SIrange{60}{70}{\celsius} followed by paraffin for \SI{12}{\hour}. The stained tissues were cut into \SI{7}{\micro\meter} slices, dewaxed, mounted with neutral balsam and then imaged using Nikon NIS-Elements microscopy.
        
\subsection*{Search Engines Methods}
\label{sup:search_engine_methods}

In the \textbf{Yottixel} method, the initial preprocessing step involves segmenting the foreground from the background in large whole slide images (WSIs). The segmented foreground is then divided into patches of size $1000 \times 1000$ for $20\times$ slides and $2000 \times 2000$ for $40\times$ slides. The $2000 \times 2000$ patches are resized to $1000 \times 1000$ before being input to the feature extractor. These patches undergo clustering using the K-means algorithm, resulting in 9 clusters based on the RGB histogram of each patch. A further selection process is applied, retaining 15\% of the patches in each cluster using another K-means clustering method based on the spatial coordinates. This final collection of patches forms a "mosaic." The Yottixel model, as recommended by its authors \parencite{kalra_yottixel_2020}, employs KimiaNet \parencite{riasatian_fine-tuning_2021}, a fine-tuned version of DenseNet specifically designed for histopathology slides, as the primary feature extractor (\cref{fig:summary_of_methods}a). The outputs of the feature extractor undergo barcoding, where binary codes are generated from the extracted features. Thus, each WSI is represented by a set of barcodes (BoBs). The database comprises BoBs for each slide in the dataset. The distance between two BoBs is calculated as the median of the minimum Hamming distances \parencite{hamming_error_1950} between each barcode in the first BoB and all barcodes in the second BoB. When a query slide is introduced, it is converted into a BoB. The distance between the query BoB and all BoBs in the database is computed, and the top 5 slides with the lowest distances are returned. For patch retrieval, the query BoB is not required, and instead, the top 5 patches from all BoBs with the minimum distances to the query patch are retrieved.

        The \textbf{SISH} method uses a similar approach to Yottixel for mosaic generation, with patch sizes of $1024 \times 1024$ for $20\times$ slides and $2048 \times 2048$ for $40\times$ slides. After mosaic generation, artifacts such as pure white patches are filtered out. The feature extraction in SISH consists of two parts: feature and index (\cref{fig:summary_of_methods}b). The feature extraction process is the same as Yottixel, where each patch in the mosaic is fed into a pretrained DenseNet, and the resulting features are binarized. The index, however, is obtained from a pretrained VQ-VAE. The patch is encoded, resulting in a latent code, which is then subjected to three layers of average pooling. The output of these layers is multiplied by scaling factors, and the sum of these results represents the index in the VEB tree. This creates the database. When querying a slide, it is converted into a mosaic, and indices and features are generated from the patches in the query mosaic. The "guided VEB search" algorithm is utilized, leveraging the properties of VEB trees, forward and backward searches, and entropy-based uncertainty calculations to retrieve the top slides based on hamming distance. The ranking algorithm accounts for class imbalance when returning the results. For patch retrieval, an index and feature are created for the query patch using a similar approach, and the best matches are found among the patches in the mosaics of the database. They also use a hamming distance threshold of $128$ to make sure they only keep high quality results. That is why sometimes they return only a few matches. 

        \textbf{RetCCL}, drawing inspiration from both Yottixel and SISH, adopts a distinct approach. Instead of clustering patches based on RGB histogram values, RetCCL first obtains contrastive-based feature vectors for each patch within the segmented foreground tiles. These features serve as inputs for a 9-class K-means clustering. Within each cluster, an additional K-means process based on spatial coordinates is performed to select 20\% of the patches. These selected patches form the mosaics, which constitute the database. The proposed feature extraction algorithm in RetCCL utilizes a clustering-guided contrastive learning method, employing the InfoNCE loss introduced in \parencite{oord_representation_2019} (\cref{fig:summary_of_methods}c). Given the prevalence of normal cells in WSIs, learning irregularities from a limited number of patches becomes crucial. The self-supervised feature extractor employs two InfoNCE losses to capture irregular regions in patches.  It worth mentioning that RetCCL was not able to perform the indexing on UPENN GBM dataset within a reasonable run time and was not utilized for certain experiments involving this dataset. 

        For retrieving similar slides, a query slide is first transformed into a mosaic, generating a set of features for each patch in the mosaic. Similarity between two patches is measured using cosine similarity between their feature vectors. The retrieval process involves returning a set of patches in the database that exhibit a similarity score of at least 70\% to the query patch. Each query patch and its corresponding results form a "bag." To account for class imbalance, an entropy-based uncertainty measure is calculated based on the occurrence of each label within the bag. Patch members in the bag are sorted according to this entropy measure. A threshold is then determined to remove lower quality results. Ultimately, the top 5 samples within each bag are returned as the final results for slide retrieval. For patch retrieval, only the top 5 patches with the highest cosine similarity scores are returned.

        In \textbf{HSHR}, the first step is to train the ResNet18 \parencite{he_deep_2015} backbone encoder using the SimCLR \parencite{chen_simple_2020} approach. Unlike the other methods, authors had not provided their backbone pre-trained weights, so we trained it from the scratch. The training data for this backbone was approximately $ 508 \times 100 = 50,800$ randomly patches of size $224 \times 224$. We trained it for 200 epochs using the same hyperparameters as the authors on 2 Nvidia A100 GPUs. Once the backbone is trained, it is used to extract the features for all densely-patched patches for each WSI. These features are used to train a 20-class k-means clustering algorithm. The features of the centroids of these clusters create the mosaic for each WSI. These features are then passed to CaEncoder for generating the hashes and attention weights. Unlike the backbone, the authors had provided the weights for their CaEncoder, and we used the same weights in our experiments. The outputs of CaEncoder is further used to create a hypergraph for the database using Eq. 15 in \parencite{li_high-order_2023}. We used $K=10$ in this equation.

        Once the hypergraph for database is constructed, each query slide would go through the same pipeline and becomes and turns into hash codes and attention weights. Using these values, the query can be appended to the hypergraph as a new vertex and a new hyperedge. Updating the hypergraph, the similarity score between this vertex and all vertices in the database can be calculate using Eq. 19 in \parencite{li_high-order_2023}. Then the top-k results are returned. We used $\alpha = \beta = 1$ in this equation.

\subsubsection*{Methodological Breakdown of Query and Retrieval Processes}

\begin{algorithm}[H]
\label{alg1}
\caption{Yottixel Algorithm}
\KwIn{Image $I$}
\KwOut{Top 5 retrieved slides}
Patch the image densely to get patches $p_1, p_2, \ldots$\;
Perform RGB histogram clustering\;
Perform spatial clustering and calculate mosaic patches\;
Feed mosaic patches to KimiaNet for feature extraction and calculate barcode for each patch\;
\ForEach{patch in input slide}{
    Calculate hamming distance between barcode of input patch with barcodes of patches from all slides in database\;
}
Choose the median of the list of minimum hamming distances for each slide in database\;
Retrieve the slides with the top five smallest medians\;
\end{algorithm}

\begin{algorithm}[H]
\label{alg2}
\caption{SISH Algorithm}
\KwIn{Image $I$}
\KwOut{Similar slides to query slide $I$}
Patch the image densely to get patches $p_1, p_2, \ldots$\;
Perform RGB histogram clustering\;
Perform spatial clustering and create mosaic patches\;
Feed mosaic patches to DenseNet and VQ-VAE encoder to calculate parameters $h$ and $m$ respectively, and create the VEB tree\;
Apply guided-search algorithm to tuples of $m$ and $h$ to calculate corresponding set of tuples $r$\;
Create a set of candidate indices $mi,c+$ and $mi,c-$ along with the original $mi$\;
Call helper functions forward-search and backward-search on $mi,c+$ and $mi,c-$ respectively\;
Take the results $RI = \{r_1, r_2, \ldots, r_k\}$ from Guided-Search as input by Results Ranking Algorithm\;
Return similar slides to query slide $I$\;
\end{algorithm}

\begin{algorithm}[H]
\label{alg3}
\caption{RetCCL Algorithm}
\KwIn{Image $I$}
\KwOut{Top $k$ similar WSIs}
Patch the image densely to get patches $p_1, p_2, \ldots$\;
Feed patches to feature extraction algorithm\;
Cluster based on extracted features, then on coordinates to create mosaic patches\;
\ForEach{patch in query WSI}{
    Perform Knn search to retrieve a bag of most similar patches in database to each patch, using cosine similarity in pretrained SSL encoder's learned embedding space\;
}
Calculate entropy within each bag, reorder bags by entropy\;
Remove bags with low quality based on mean of cosine similarity scores in top-5\;
\ForEach{bag}{
    Perform voting for each diagnosis within the bag, get the top-5 samples, then do majority vote to get associated WSI\;
}
Retrieve top-$k$ similar WSIs\;
\end{algorithm}

\begin{algorithm}[H]
\label{alg4}
\caption{HSHR Algorithm}
\KwIn{Image $I$}
\KwOut{Top $k$ similar WSIs}
Patch the image densely to get patches $p_1, p_2, \ldots$\;
Feed patches to feature extraction algorithm\;
Cluster based on extracted features, then on coordinates to create mosaic patches\;
\ForEach{patch in query WSI}{
    Create a bag containing the query patch and its retrieved patches\;
}
Calculate entropy within each bag, reorder bags by entropy\;
Remove bags with low quality based on mean of cosine similarity scores in top-5\;
\ForEach{bag}{
    Perform voting for each diagnosis within the bag, get the top-5 samples, then do majority vote to get associated WSI\;
}
Retrieve top-$k$ similar WSIs\;
\end{algorithm}

\subsubsection*{Complexity Analysis}

The time complexities of the key operations in different algorithms are summarized in Table \cref{tab:complexity_analysis}. In the table, $n$ (Yottixel) denotes the length of the hamming vector, $m$ (Yottixel) denotes the maximum number of patches for each WSI, $T$ (common across Yottixel, HSHR, and RetCCL) denotes the total number of slides in the database, $B$ (SISH) denotes the number of patches in a WSI, $K$ (RetCCL, HSHR) denotes the number of patches of the query slide, $M$ (RetCCL, HSHR) denotes the total number of diagnoses in the database, and $H$ (HSHR) denotes the hash vector size.

\begin{stable}[ht]
    \centering
    \caption{Time complexity analysis for different components of the compared methods.}
    \label{tab:complexity_analysis}
    \begin{tabular}{p{2cm}p{7cm}}
        \toprule
        \textbf{Algorithm} & \textbf{Operation and Time Complexity} \\
        \midrule
        \multirow{3}{*}{\textbf{Yottixel}} & Hamming distance calculation: $O(n.T.m^2)$ \\
        & Minimum: $O(T.m^2)$ \\
        & Median: $O(T.\log(T))$ \\
        \midrule
        \multirow{2}{*}{\textbf{SISH}} & Search Performance: $O(1)$ \\
        & Ranking: $O(B’)$, $B’ = 0.05 \cdot B$ \\
        \midrule
        \multirow{6}{*}{\textbf{RetCCL}} & Cosine similarity calculation: $O(K.B)$ \\
        & Probability calculation: $O(M.K.B)$ \\
        & Entropy calculation: $O(K.M)$ \\
        & Sorting bags based on entropies: $O(K^2)$ \\
        & Mean of cosine similarity scores: $O(T.K)$ \\
        & Removing bags with low quality: $O(K)$ \\
        \midrule
        \multirow{8}{*}{\textbf{HSHR}} & Hamming distance: $O(H.T.M^2)$ \\
        & Sorting: $O(M)$ \\
        & Incidence Matrix calculation: $O(K.M.T)$ \\
        & Cross similarity: $O(T^2)$ \\
        & Vertex similarity: $O(T^3)$ \\
        & Hyperedge similarity: $O(T^3)$ \\
        & Final sorting: $O(T)$ \\
        \bottomrule
    \end{tabular}
\end{stable}

\subsection*{Studies}

\subsubsection*{Performance Metrics}
As discussed, we use majority voting and average precision at $k$ as main performance metrics in different experiments. Supplementary Algorithms \cref{alg5} and \cref{alg6} summarize they way we defined these metrics. The important point about majority voting at $k$ is that if it returns \texttt{None}, that sample would not be counted in the average, while 0 outputs are counted towards average. 

\begin{algorithm}[H]
\label{alg5}
\DontPrintSemicolon
\caption{Majority Voting at k}

\KwData{$row$, $k$}
\KwResult{Result of the majority vote}

$votes \gets \text{empty list}$\;

\For{$i = 1$ \KwTo $k$}{
    $ret\_site \gets row[\text{{f'ret\_i\_site'}}]$\;
    \eIf{$ret\_site$ is null}{
        append $-1$ to $votes$\;
    }{
        append $ret\_site$ to $votes$\;
    }
}

\If{not $votes$}{
    \Return 0\;
}

$counter \gets \text{Counter}(votes)$\;
$most\_common \gets counter.\text{most\_common}(1)$\;

\eIf{$most\_common[0][0] = row[\text{{'query\_site'}}]$}{
    \Return 1\;
}{
    \eIf{$most\_common[0][0] = -1$}{
        \Return None\;
    }{
        \Return 0\;
    }
}
\end{algorithm}

\begin{algorithm}[H]
\label{alg6}
\DontPrintSemicolon
\caption{Average Precision at k}

\KwData{$row$, $k$}
\KwResult{Average precision at k}

$relevant\_count \gets 0$\;
$precision\_sum \gets 0$\;

\For{$i = 1$ \KwTo $k$}{
    $ret\_site \gets row[\text{{f'ret\_i\_site'}}]$\;
    \If{$ret\_site$ is None}{
        \textbf{continue}\;
    }
    \If{$ret\_site = row[\text{{'query\_site'}}]$}{
        $relevant\_count \gets relevant\_count + 1$\;
        $precision\_sum \gets precision\_sum + \frac{relevant\_count}{i}$\;
    }
}

\If{$relevant\_count = 0$}{
    \Return 0\;
}
\Return $\frac{precision\_sum}{\text{min}(k, relevant\_count)}$\;
\end{algorithm}

\subsubsection*{Reader study}
Seven pathologists were shown the top three ranked results on patch-level retrieval for one patch across three different H\&E slides. All pathologists were shown the original queried patch of interest but were blinded to the algorithms involved, ranked order of the retrieved patches, and diagnoses of both the queried and retrieved patched. Pathologists were then asked for their Mean Opinion Score (MOS) based on their perspective on the quality of the results, ranked from one to five with higher scores indicating higher quality.

            \begin{sfigure}[!ht]
                \centering
                \includegraphics[width=\linewidth]{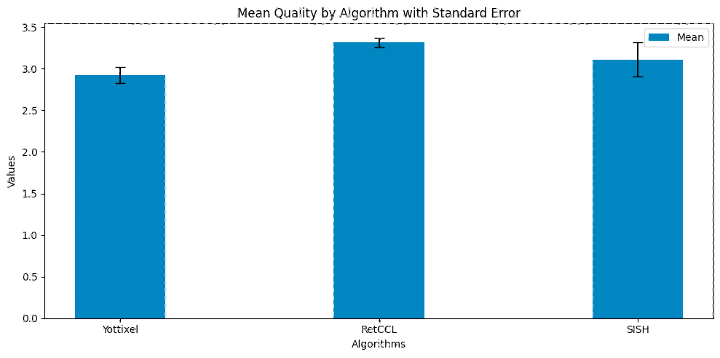}
                \caption{Reader Study: Average quality ratings and standard deviation for each algorithm as evaluated by seven pathologists.}

                \label{fig:reader_study}
            \end{sfigure}

        \begin{stable}[!ht]
            \centering
            \caption{Evaluation results of different methods on Reader Study slides for primary site retrieval task.}
            \label{suptab:primary_site_evaluation_reader}
            \begin{tabular}{p{2cm}p{3.3cm}p{1.2cm}p{1.2cm}p{1.2cm}p{1.2cm}p{1.2cm}p{1.2cm}}
                \toprule
                \textbf{Method} & \textbf{Reader Study} & \textbf{MV@1} & \textbf{MV@3} & \textbf{MV@5} & \textbf{MV@10} & \textbf{AP@3} & \textbf{AP@5} \\
                \midrule
                \multirow{3}{*}{\shortstack{\textbf{YOTTIXEL} \\ \textbf{+ KimiaNet}}} & \textbf{MSB-09151-01-11} & 0 & 0 & 0 & 0 & 0 & 0.250 \\
                & \textbf{MSB-09977-01-22} & 1 & 1 & 1 & 0 & 1 & 1 \\
                & \textbf{Her2Pos\_Case\_66} & 1 & 1 & 1 & 1 & 0.833 & 0.806 \\
                \midrule
                \multirow{3}{*}{\shortstack{\textbf{SISH +} \\ \textbf{DenseNet}}} & \textbf{MSB-09151-01-11} & 0 & 0 & 0 & 0 & 0 & 0 \\
                & \textbf{MSB-09977-01-22} & 0 & 0 & 0 & 0 & 0 & 0 \\
                & \textbf{Her2Pos\_Case\_66} & 0 & 0 & 0 & 0 & 0 & 0 \\
                \midrule
                \multirow{3}{*}{\textbf{RetCCL}} & \textbf{MSB-09151-01-11} & 0 & 0 & 0 & 1 & 0 & 0.250 \\
                & \textbf{MSB-09977-01-22} & 1 & 1 & 0 & 0 & 0 & 1 \\
                & \textbf{Her2Pos\_Case\_66} & 0 & 0 & 0 & 1 & 0 & 0.325 \\
                \midrule
                \multirow{3}{*}{\textbf{HSHR}} & \textbf{MSB-09151-01-11} & 1 & 1 & 1 & 1 & 0.833 & 0.833 \\
                & \textbf{MSB-09977-01-22} & 1 & 1 & 1 & 1 & 1 & 0.700 \\
                & \textbf{Her2Pos\_Case\_66} & 1 & 1 & 1 & 0 & 1 & 0.867 \\
                \bottomrule
            \end{tabular}
        \end{stable}

        \begin{stable}[!ht]
            \centering
            \caption{Evaluation results of different methods on Reader Study slides for subtype retrieval task.}
            \label{suptab:subtype_evaluation_reader}
            \begin{tabular}{p{2.5cm}p{3.5cm}p{1.2cm}p{1.2cm}p{1.2cm}p{1.2cm}p{1.2cm}}
                \toprule
                \textbf{Method} & \textbf{Reader Study} & \textbf{MV@1} & \textbf{MV@3} & \textbf{MV@5} & \textbf{AP@3} & \textbf{AP@5} \\
                \midrule
                \multirow{3}{*}{\shortstack{\textbf{YOTTIXEL} \\ \textbf{+ KimiaNet}}} & \textbf{MSB-09151-01-11} & - & - & - & - & - \\
                & \textbf{MSB-09977-01-22} & 0 & 0 & 0 & 0 & 0 \\
                & \textbf{Her2Pos\_Case\_66} & - & - & - & - & - \\
                \midrule
                \multirow{3}{*}{\shortstack{\textbf{SISH +} \\ \textbf{DenseNet}}} & \textbf{MSB-09151-01-11} & - & - & - & - & - \\
                & \textbf{MSB-09977-01-22} & 0 & 0 & 0 & 0 & 0 \\
                & \textbf{Her2Pos\_Case\_66} & - & - & - & - & - \\
                \midrule
                \multirow{3}{*}{\textbf{RetCCL}} & \textbf{MSB-09151-01-11} & - & - & - & - & - \\
                & \textbf{MSB-09977-01-22} & 0 & 1 & 1 & 0.583 & 0.589 \\
                & \textbf{Her2Pos\_Case\_66} & - & - & - & - & - \\
                \midrule
                \multirow{3}{*}{\textbf{HSHR}} & \textbf{MSB-09151-01-11} & - & - & - & - & - \\
                & \textbf{MSB-09977-01-22} & 1 & 1 & 1 & 0.833 & 0.806 \\
                & \textbf{Her2Pos\_Case\_66} & - & - & - & - & - \\
                \bottomrule
            \end{tabular}
        \end{stable}

\begin{stable}[!h]
\centering
\caption{Mann-Whitney U Test Results for Microscope}
\label{tab:u_test_results_microscope}
\begin{tabular}{lccr}
\toprule
\textbf{Microscope} & \textbf{Metric} & \textbf{U-statistic} & \textbf{P-value} \\
\midrule
\multirow{4}{*}{SISH} 
& MV\_at\_10\_site & nan & nan \\
& AP\_at\_5\_site & 578.000 & 1.0000000000000000 \\
& MV\_at\_5\_subtype & nan & nan \\
& AP\_at\_5\_subtype & 543.000 & 0.6156890923465373 \\
\midrule
\multirow{4}{*}{HSHR} 
& MV\_at\_10\_site & 658.500 & 0.0670269674521155 \\
& AP\_at\_5\_site & 523.500 & 0.6135535212442216 \\
& MV\_at\_5\_subtype & 672.500 & 0.0905435676918205 \\
& AP\_at\_5\_subtype & 636.000 & 0.3353506099579945 \\
\midrule
\multirow{4}{*}{Yottixel} 
& MV\_at\_10\_site & 523.500 & 0.5298120557113710 \\
& AP\_at\_5\_site & 455.500 & 0.1566946582170631 \\
& MV\_at\_5\_subtype & 532.000 & 0.6415761864226559 \\
& AP\_at\_5\_subtype & 534.500 & 0.7372510705087723 \\
\bottomrule
\end{tabular}
\end{stable}

\begin{stable}[!h]
\centering
\caption{Mann-Whitney U For Her2+}
\label{tab:u_test_results}
\begin{tabular}{lccr}
\toprule
\textbf{Model} & \textbf{Metric} & \textbf{U-statistic} & \textbf{P-value} \\
\midrule
\multirow{2}{*}{Yottixel} & MV\_at\_10\_site & 2906.500 & 0.0000006008842402 \\
& AP\_at\_5\_site & 3287.500 & 0.0008328526912513 \\
\midrule
\multirow{2}{*}{SISH} & MV\_at\_10\_site & nan & nan \\
& AP\_at\_5\_site & 3651.500 & 0.0130448102685896 \\
\midrule
\multirow{2}{*}{RetCCL} & MV\_at\_10\_site & nan & nan \\
& AP\_at\_5\_site & 2483.500 & 0.0000000486711422 \\
\midrule
\multirow{2}{*}{HSHR} & MV\_at\_10\_site & 3407.500 & 0.0002646668324846 \\
& AP\_at\_5\_site & 3274.500 & 0.0008856874714843 \\
\bottomrule
\end{tabular}
\end{stable}

\begin{sfigure}[!ht]
    \centering
    \includegraphics[width=\linewidth]{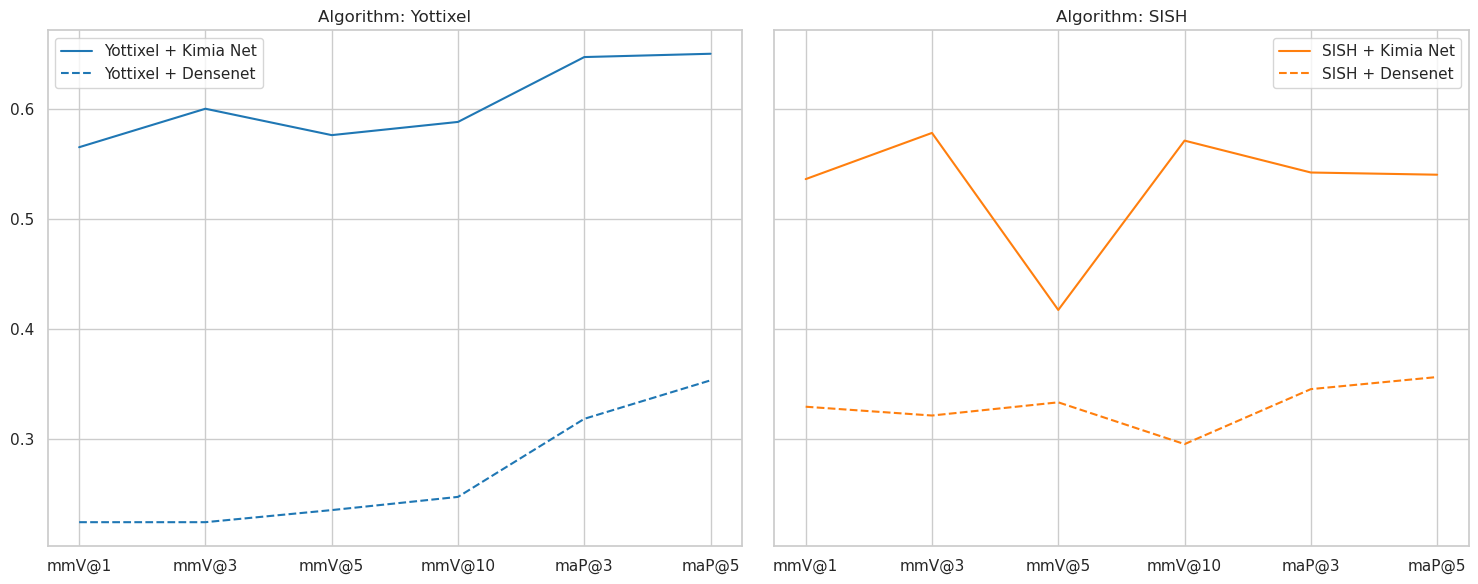}
    \caption{Performance comparison of Yottixel and SISH algorithms using Kima Net and Densenet on the Yale Trastuzumab dataset across various metrics.}

    \label{fig:algorithm_comparison}
\end{sfigure}

\begin{sfigure}[!ht]
    \centering
    \includegraphics[width=\linewidth]{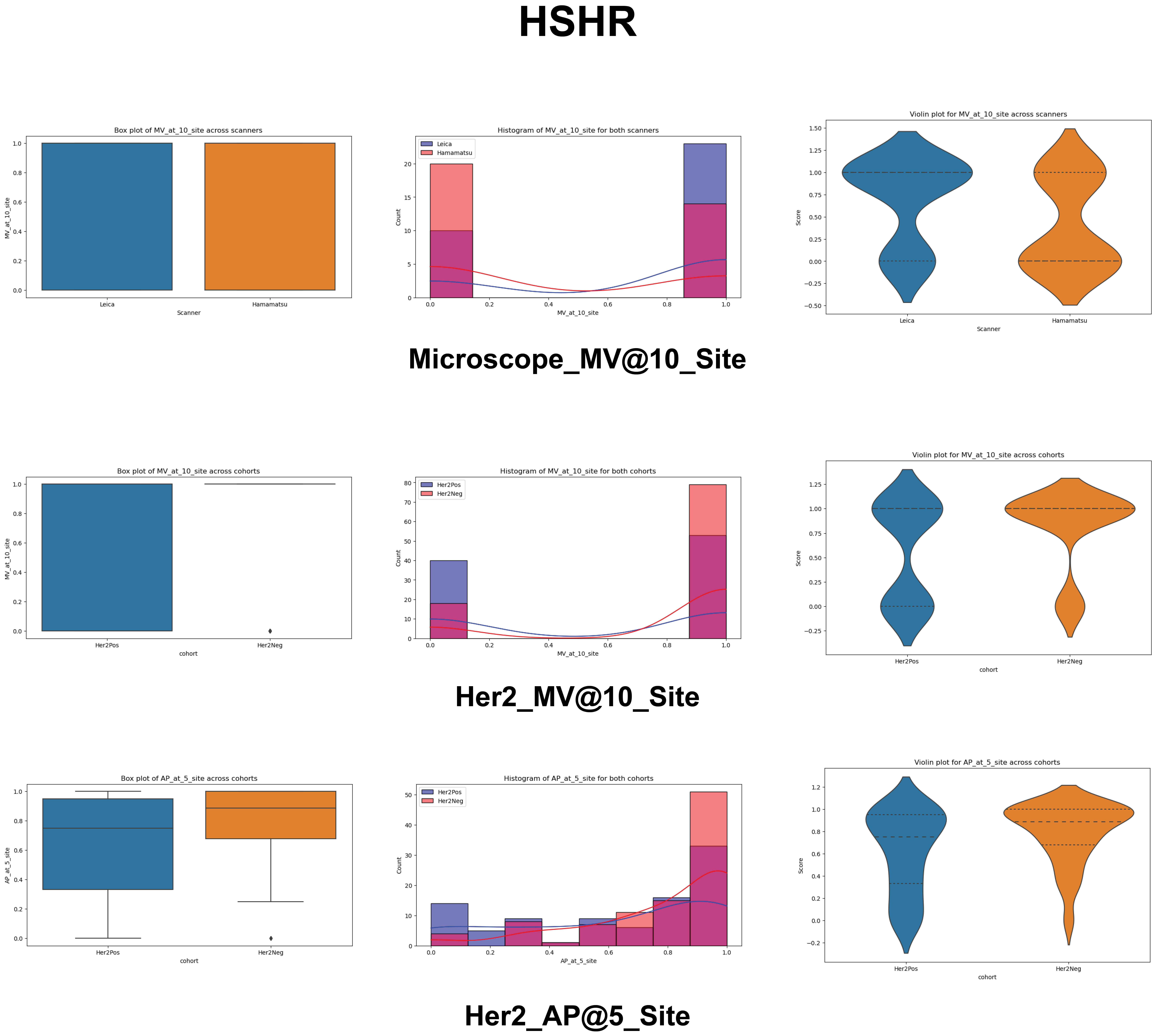}
    \caption{Metrics for HSHR Microscope comparison analysis and Her2 analysis.}

    \label{fig:algorithm_comparison}
\end{sfigure}

\begin{sfigure}[!ht]
    \centering
    \includegraphics[width=\linewidth]{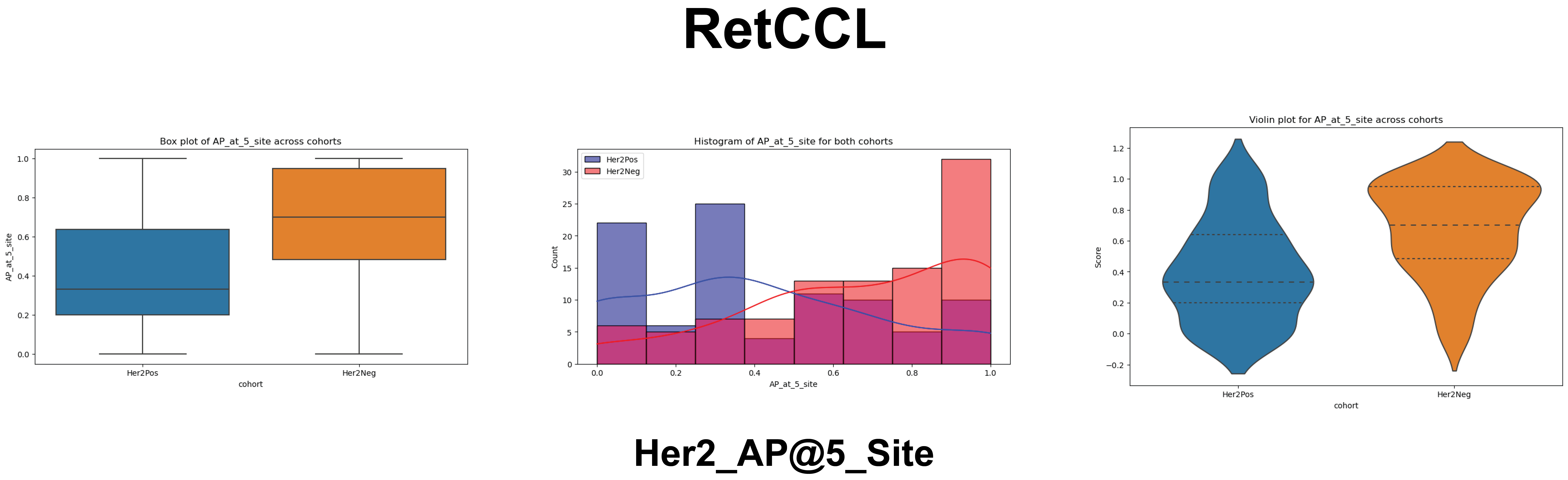}
    \caption{Metrics for RetCCL Her2 analysis.}

    \label{fig:algorithm_comparison}
\end{sfigure}

\begin{sfigure}[!ht]
    \centering
    \includegraphics[width=\linewidth]{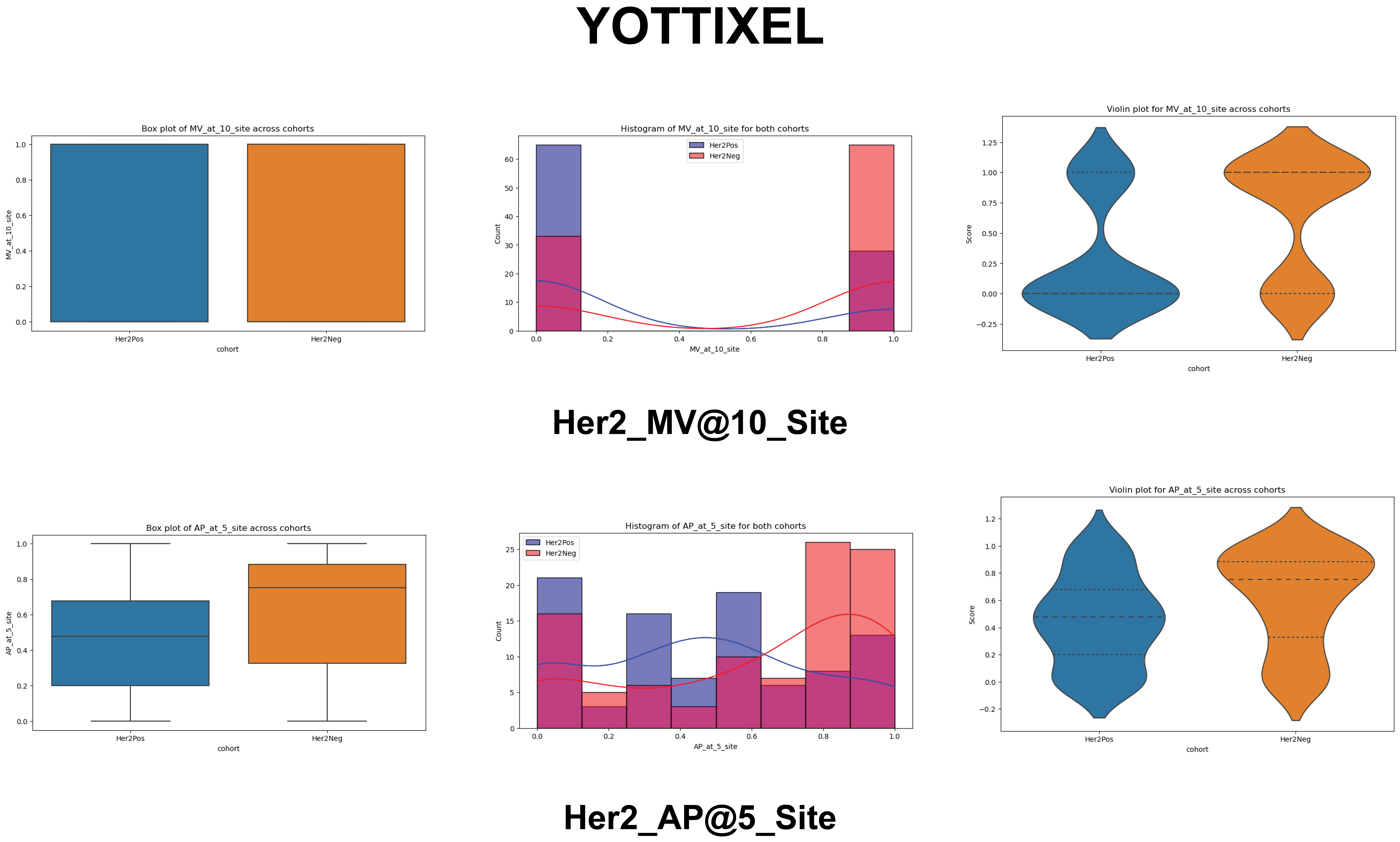}
    \caption{Metrics for YOTTIXEL Her2 analysis.}

    \label{fig:algorithm_comparison}
\end{sfigure}

\begin{sfigure}[!ht]
    \centering
    \includegraphics[width=\linewidth]{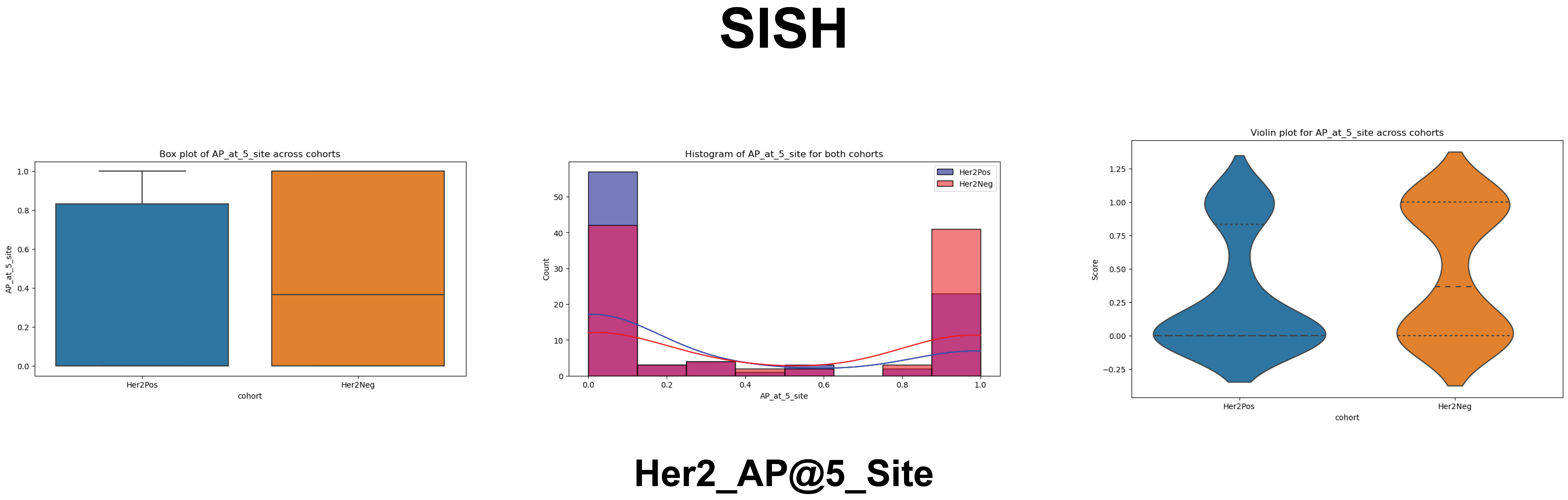}
    \caption{Metrics for SISH Her2 analysis.}

    \label{fig:algorithm_comparison}
\end{sfigure}

\subsection*{Extended Results}

        We choose three internal patient cases for visual review of tissue, subtype, and patch-level search results. Slide 1 is a Lung Adenocarcinoma (LUAD) case from a patient's partial lobectomy. Slide 2 is a Low Grade Glioma (LGG) that was retrieved by surgical biopsy. Slide 3  is from a patient with Hepatocellular Carcinoma (LIHC) who underwent a liver biopsy by fine needle aspiration. 

        \begin{stable}[!ht]
                \centering
                \caption{Evaluation results of different methods on large datasets for primary site retrieval task. \textit{mMV} means mean Majority Voting score and mAP means mean Average Precision.}
                \label{suptab:primary_site_evaluation_rest}
                \begin{tabular}{p{2cm}p{3.3cm}p{1.5cm}p{1.5cm}p{1.5cm}p{1.5cm}p{1.5cm}p{1.5cm}}
                    \toprule
                    \textbf{Method} & \textbf{Dataset} & \textbf{mMV@1} & \textbf{mMV@3} & \textbf{mMV@5} & \textbf{mMV@10} & \textbf{mAP@3} & \textbf{mAP@5} \\
                    \midrule
                    \multirow{5}{*}{\shortstack{\textbf{YOTTIXEL} \\ \textbf{+ KimiaNet}}} & \textbf{UPENN GBM} & 0.706 & 0.735 & 0.735 & 0.794 & 0.752 & 0.765 \\
                    & \textbf{CPTAC GBM} & 0.606 & 0.606 & 0.667 & 0.727 & 0.646 & 0.644 \\
                    & \textbf{Yale Her2 Pos} & 0.548 & 0.559 & 0.538 & 0.570 & 0.627 & 0.624 \\
                    & \textbf{Yale Her2 Neg} & 0.742 & 0.742 & 0.784 & 0.814 & 0.809 & 0.788 \\
                    & \textbf{Yale Trastuzumab} & 0.565 & 0.600 & 0.576 & 0.588 & 0.647 & 0.650 \\
                    \midrule
                    \multirow{5}{*}{\shortstack{\textbf{SISH +} \\ \textbf{DenseNet}}} & \textbf{UPENN GBM} & 0.735 & 0.706 & 0.706 & 0.758 & 0.730 & 0.720 \\
                    & \textbf{CPTAC GBM} & 0.706 & 0.697 & 0.697 & 0.710 & 0.725 & 0.729 \\
                    & \textbf{Yale Her2 Pos} & 0.269 & 0.247 & 0.221 & 0.167 & 0.294 & 0.304 \\
                    & \textbf{Yale Her2 Neg} & 0.454 & 0.442 & 0.446 & 0.375 & 0.476 & 0.480 \\
                    & \textbf{Yale Trastuzumab} & 0.329 & 0.321 & 0.333 & 0.295 & 0.345 & 0.356 \\
                    \midrule
                    \multirow{5}{*}{\textbf{RetCCL}} & 
                    \textbf{UPENN GBM} & - & - & - & - & - & - \\
                    & \textbf{CPTAC GBM} & 0.588 & 0.594 & 0.750 & 0.846 & 0.728 & 0.736 \\
                    & \textbf{Yale Her2 Pos} & 0.172 & 0.301 & 0.355 & 0.481 & 0.345 & 0.404 \\
                    & \textbf{Yale Her2 Neg} & 0.510 & 0.602 & 0.742 & 0.864 & 0.656 & 0.671 \\
                    & \textbf{Yale Trastuzumab} & 0.212 & 0.247 & 0.353 & 0.420 & 0.382 & 0.428 \\
                    \midrule
                    \multirow{5}{*}{\textbf{HSHR}} & 
                    \textbf{UPENN GBM} & 0.794 & 0.824 & 0.765 & 0.735 & 0.816 & 0.809 \\
                    & \textbf{CPTAC GBM} & 0.758 & 0.727 & 0.818 & 0.909 & 0.813 & 0.810 \\
                    & \textbf{Yale Her2 Pos} & 0.247 & 0.323 & 0.323 & 0.301 & 0.429 & 0.449 \\
                    & \textbf{Yale Her2 Neg} & 0.571 & 0.592 & 0.622 & 0.663 & 0.632 & 0.612 \\
                    & \textbf{} & 0.447 & 0.447 & 0.494 & 0.529 & 0.581 & 0.569 \\
                    \midrule
                    \multirow{2}{*}{\shortstack{\textbf{YOTTIXEL} \\ \textbf{+ DenseNet}}} & \textbf{Yale Trastuzumab} & 0.224 & 0.224 & 0.235 & 0.247 & 0.318 & 0.353 \\
                    & & & & &\\
                    \midrule
                    \multirow{2}{*}{\shortstack{\textbf{SISH} \\ \textbf{+ KimiaNet}}} & \textbf{Yale Trastuzumab} & 0.536 & 0.578 & 0.417 & 0.571 & 0.542 & 0.540 \\ 
                    & & & & &\\
                    \bottomrule
                \end{tabular}
            \end{stable}

            \begin{stable}[!ht]
                \centering
                \caption{Evaluation results of different methods on large datasets for subtype retrieval task.}
                \label{suptab:primary_subtype_evaluation_rest}
                \begin{tabular}{p{2cm}p{3.3cm}p{1.5cm}p{1.5cm}p{1.5cm}p{1.5cm}p{1.5cm}}
                    \toprule
                    \textbf{Method} & \textbf{Dataset} & \textbf{mMV@1} & \textbf{mMV@3} & \textbf{mMV@5} & \textbf{mAP@3} & \textbf{mAP@5} \\
                    \midrule
                    \multirow{2}{*}{\shortstack{\textbf{YOTTIXEL} \\ \textbf{+ KimiaNet}}} & \textbf{UPENN GBM} & 0.382 & 0.382 & 0.294 & 0.468 & 0.454 \\
                    & \textbf{CPTAC GBM} & 0.303 & 0.303 & 0.242 & 0.399 & 0.418 \\
                    \midrule
                    \multirow{2}{*}{\shortstack{\textbf{SISH +} \\ \textbf{DenseNet}}} & \textbf{UPENN GBM} & 0.706 & 0.656 & 0.677 & 0.725 & 0.725 \\
                    & \textbf{CPTAC GBM} & 0.676 & 0.667 & 0.656 & 0.686 & 0.681 \\
                    \midrule
                    \multirow{2}{*}{\textbf{RetCCL}} & 
                    \textbf{UPENN GBM} & - & - & - & - & - \\
                    & \textbf{CPTAC GBM} & 0.529 & 0.375 & 0.406 & 0.667 & 0.648 \\
                    \midrule
                    \multirow{2}{*}{\textbf{HSHR}} & 
                    \textbf{UPENN GBM} & 0.618 & 0.559 & 0.559 & 0.721 & 0.704 \\
                    & \textbf{CPTAC GBM} & 0.758 & 0.727 & 0.758 & 0.806 & 0.785 \\
                    \bottomrule
                \end{tabular}
            \end{stable}

\end{document}